\documentclass[aps,prd,reprint,preprintnumbers,groupedaddress,nofootinbib]{revtex4-1}
\def\ea {\varepsilon _\bot p_h}
\def\eb { \frac{ \varepsilon _\bot k}R}
\usepackage{graphics} 
\usepackage{amsbsy}

\begin{document}

\preprint{JLAB-PHY-19-2938}

\title{Lowest order QED radiative effects in polarized SIDIS}


\author{Igor Akushevich}
\email[]{igor.akushevich@duke.edu}
\altaffiliation{Jefferson Lab., Newport News, Virginia 23606, USA}
\affiliation{Physics Department, Duke University, Durham, Noth Carolina 27708, USA}
\author{Alexander Ilyichev}
\email[]{ily@hep.by}
\affiliation{Research Institute for Nuclear Problems Byelorussian State University, Minsk, 220030, Belarus}

\date{\today}

\begin{abstract}
The explicit exact analytical expressions for the lowest order radiative corrections to the semi-inclusive deep inelastic scattering of the polarized particles are obtained in the most compact, covariant form 
convenient for the numerical analysis. The infrared divergence from the real photon emission 
is extracted and canceled using the Bardin-Shumeiko approach. The contribution of the exclusive radiative tail is presented. 
The analytic results obtained within the ultrarelativistic approximation are also shown.  
\end{abstract}


\maketitle
\section{\label{Intro}Introduction}
Nowadays the polarized semi-inclusive deep-inelastic scattering (SIDIS)  plays a crucial role in our 
understanding of the internal spin structure of the nucleons. Information
on the three-dimensional structure of the polarized proton and neutron can be obtained 
by extracting the quark transverse
momentum distributions from the various single spin asymmetries measured in SIDIS with polarized particles. 
Specifically, the Sivers and Collins contributions can be selected \cite{Barone} from the present data on transversely polarized targets ${\vec p}(e, e^\prime \pi )x$ in HERMES
\cite{HERMES}, ${\vec D}(\mu, \mu^\prime \pi )x$ in COMPASS
\cite{COMPASS}  and  $^3{\vec {\rm He}}(e, e^\prime \pi )x$ in JLab \cite{JLab1} which show 
a strong flavor dependence of transverse
momentum distributions. 
Moreover in the near future, highly accurate experiments are planned at 12 GeV Jlab~\cite{12Jlab} 
that will provide unique opportunities for the breakthrough in the investigation of the  nucleon 
 structure by carrying out multidimensional precision studies of
longitudinal and transverse spin and momentum degrees of freedom from SIDIS experiments with
high luminosity in combination with large acceptance detectors.

It is well known that one of the important sources of the systematical uncertainties in SIDIS experiments with and without 
polarization of initial particles are the QED radiative corrections (RC). RC to the threefold differential 
cross section ($d\sigma/dxdydz$, where {$x$ and $y$ are the standard Bjorken variables 
and the $z$ is the fraction of the virtual-photon energy transferred to the detected hadron})
can be calculated using the patch SIRAD of FORTRAN code POLRAD \cite{Polrad} created based on the original calculations in Refs. \cite{SSh1,SSh2} for unpolarized and polarized particles. The calculation of RC to the fivefold differential cross section of unpolarized particles ($d\sigma/dxdydzdp_t^2d\phi_h $, where $p_t$ is the detected hadron  
transverse momentum and  $\phi_h$ is the azimuthal angle
between the lepton scattering and hadron production  planes)
 was performed in Ref. \cite{ASSh}. These calculations did not contain the radiative tail from the exclusive reactions as a separate contribution involving the exclusive structure functions (SF). This limitation
was addressed in Ref. \cite{AIO} in which the authors explicitly calculated the exclusive radiative tail and implemented the exclusive SF using the approach of MAID \cite{Maid}.

In the present paper we consider the general task of RC calculation when the initial nucleon can be arbitrarily polarized. The analytical expressions for RC to SIDIS are obtained
for the sixfold cross section  with the longitudinally polarized lepton and arbitrarily polarized 
target, $d\sigma/dxdydz dp_t^2d\phi_h d\phi$, where {the azimuthal angle $\phi$
between the lepton scattering and ground planes is introduced to appropriately account for the transverse target polarization}.
The contribution of the exclusive radiative tail to the total RC is also
presented. Similar to the previous analyses we calculated RC in the model-independent way. These corrections are 
induced by the unobservable real photon emission from the lepton leg, leptonic vertex correction, and vacuum polarization.
The model-independent correction is proportional to the leading logarithm $\log(Q^2/m^2)$, which is large because of high transferring momentum squared $Q^2$ ($>1$~GeV$^2$) and small electron mass $m$. 
What is not accounted for in this approach is the correction due to the real and additional
virtual photon emission by hadrons including the two-photon exchange and QED hadronic vertex correction. 
However, this correction should not be accounted for in the majority of cases, e.g., when the used model for SF was extracted from the experiment in which emission by hadrons had not been applied in the RC procedure of experimental data. 

The Bardin-Shumeiko approach 
\cite{BSh}
is used for extraction and cancellation of the infrared divergence coming from the real and virtual photon emission. In contrast to the widely used Mo-Tsai approach \cite{Mo-Tsai,Tsai} the final expression for RC within the Bardin-Shumeiko 
approach
does not depend on an artificial
parameter that is introduced in \cite{Mo-Tsai,Tsai} for separation of the photon emission on the hard and soft 
parts.

In this paper we apply an approach for decomposition of the initial nucleon and virtual photon polarization
as well as the real photon four-momentum over the respective bases (Appendix~\ref{ap0}). The polarization decomposition is used
for the hadronic tensor representation in a covariant form. The momentum decomposition is used to simplify integration over the momentum of the unobserved photon. Specifically, this allows us to essentially reduce the number of pseudoscalars occurring after the convolution of the leptonic tensors of radiative effects with the hadronic tensor 
and present the final expressions for RC in a compact, covariant form convenient for numerical analysis.
All calculations have been performed in an exact way keeping the lepton mass at all stages of the calculation. The dependence of certain terms in the exact final expressions for RC on the electron mass is quite tricky, and therefore, we analyze respective contributions in the ultrarelativistic approximation allowing for extraction of the electron mass dependence explicitly and classifying all terms in RC as leading (i.e., containing the leading logarithms), next-to-leading (i.e., independent of the electron mass), and other potentially negligible terms (i.e., the terms vanishing in the approximation of $m\rightarrow 0$). Thus the results obtained in the paper contain both exact formulas for RC and expressions in ultrarelativistic approximations allowing us to explicitly control the dependence on the electron mass. Thus, the analytic expressions for RC are valid for experiments with muons (e.g., COMPASS \cite{COMPASS}) in which the approximation of the zero lepton mass could not be appropriate.


The rest of the article is organized as follows. The hadronic tensor, different sets for the SF used in the literature, as well as the
lowest order (Born) contribution to the SIDIS process are discussed in Sec.~\ref{HT}. The calculation of the lowest order QED RC
to the obser\-vables in SIDIS as well as the explicit results for both the semi-inclusive final hadronic state and exclusive radiative tail 
contributions are presented in  Sec.~\ref{LO}. The infrared divergence in these calculations are extracted from the real photon emission with the
semi-inclusive final hadronic state
by the Bardin-Shumeiko approach \cite{BSh} and then canceled with the corresponding term from the leptonic vertex correction in such a 
way that the obtained results are free from an intermediate parameter ${\bar k}_0$. For the parametrization of the infrared and
ultraviolet divergences the dimension regularization is used. The results of analyses of the exact expressions in ultrarelativistic approximation are given in 
Sec.~\ref{URA}. Particularly we show that the double 
leading logarithms coming from the terms with the soft photon emission and the leptonic vertex correction 
cancel in their sum. 
A brief discussion and conclusion are presented in 
Sec.~\ref{Conc}. Technical details and the most cumbersome parts of the RC are presented in four appendixes. 
The bases for the decomposition of
the initial target and virtual photon polarization as well
as the real photon momentum are presented in  Appendix~\ref{ap0}. 
The explicit expressions for the real photon emission quantities are presented in  Appendix~\ref{ap1}. 
The details
of the approach for the infrared divergence extraction and cancellation
are given in Appendix~\ref{ap2}. The detailed calculations of the additional virtual particle contributions are presented  
in  Appendix~\ref{ap3}.

\section{\label{HT}Hadronic Tensor and Born Contribution}
The sixfold differential  cross section of SIDIS with polarized particles  
\begin{eqnarray}
e(k_1,\xi)+n(p, \eta)\longrightarrow e(k_2)+h(p_h)+x(p_x)
\label{lo}
\end{eqnarray}
($k_1^2=k_2^2=m^2$, $p^2=M^2$, $p_h^2=m_h^2$) where $\xi$ ($\eta$) is 
the initial lepton (nucleon)
polarized vector can be described by the following
set of variables
\begin{eqnarray}
&\displaystyle
x=-\frac{q^2}{2qp},\;
y=\frac{qp}{k_1p},\;
z=\frac{p_hp}{pq},\;
\nonumber \\
&\displaystyle
t=(q-p_h)^2,
\;
\phi_h,
\;
\phi.
\label{setvar}
\end{eqnarray}
Here $q=k_1-k_2$,
$\phi_h$ is the angle between
$({\bf k_1},{\bf k_2})$ and $({\bf q},{\bf p_h})$ planes
and $\phi$ is the angle between
$({\bf k_1},{\bf k_2})$ and the ground
planes in the target rest frame  reference system (${\bf p}=0$).

Also we use the following set of invariants:
\begin{eqnarray}
&S=2pk_1,\;
Q^2=-q^2,\;
Q_m^2=Q^2+2m^2,\;
\nonumber\\&
X=2pk_2,\;
S_x=S-X,\;
S_p=S+X,
\nonumber\\&
\displaystyle
V_{1,2}=2k_{1,2}p_h,\;
V_+=\frac 1 2(V_1+ V_2),
\nonumber\\&
\displaystyle
V_-=\frac 1 2(V_1- V_2)=\frac 1 2(m_h^2-Q^2-t),
\nonumber\\&
S^\prime=2k_1(p+q-p_h)=S-Q^2-V_1,
\nonumber\\&
X^\prime=2k_2(p+q-p_h)=X+Q^2-V_2,
\nonumber\\&
p_x^2=(p+q-p_h)^2=M^2+t+(1-z)S_x.
\nonumber\\&
\lambda_S=S^2-4M^2m^2,\;
\lambda_Y=S_x^2+4M^2Q^2,\;
\nonumber \\&
\lambda_1=Q^2(SX-M^2Q^2)-m^2\lambda_Y,\;
\lambda_m=Q^2(Q^2+4m^2),\;
\nonumber\\&
\lambda_S^\prime=S^{\prime 2}-4m^2p_x^2,\;
\lambda_X^\prime=X^{\prime 2}-4m^2p_x^2.
\label{setvar2}
\end{eqnarray}

Noninvariant variables including the energy
$p_{h0}$, longitudinal $p_l$, and transverse $p_t$ ($k_t$) three-momenta of the detected hadron
(the incoming or scattering lepton)
with respect to the virtual photon direction
in the target rest frame are expressed in terms of 
the above invariants:
\begin{eqnarray}
&&\displaystyle
p_{h0}=\frac{zS_x}{2M},\; 
\nonumber\\
&&\displaystyle
p_l
=\frac{zS^2_x-4M^2V_-}{2M\sqrt{\lambda_Y}}
=\frac{zS^2_x+2M^2(t+Q^2-m_h^2)}{2M\sqrt{\lambda_Y}},
\nonumber\\[2mm]
&&\displaystyle
p_t=\sqrt{p_{h0}^2-p_l^2-m_h^2},
\nonumber\\
&&\displaystyle
k_t=\sqrt{\frac{\lambda_1}{\lambda_Y}}.
\label{setvar3}
\end{eqnarray}

As a result the quantities $V_{1,2}$ can be written through $\cos \phi_h$
and other variables defined in Eqs.~(\ref{setvar})-(\ref{setvar3}) as
\begin{eqnarray}
V_1&=&p_{h0}\frac{S}{M}-\frac{p_l(S S_x+2M^2Q^2)}{M\sqrt{\lambda_Y}}-2 p_tk_t\cos \phi_h,
\nonumber\\
V_2&=&p_{h0}\frac{X}{M}-\frac{p_l(X S_x-2M^2Q^2)}{M\sqrt{\lambda_Y}}-2 p_tk_t\cos \phi_h.
\nonumber\\
\label{cphih}
\end{eqnarray}

The sine of $\phi_h$ is expressed as 
\begin{eqnarray}
\sin \phi_h&=&-
\frac{2\varepsilon _\bot p_h}{p_t\sqrt{\lambda_1}},
\label{sphih}
\end{eqnarray}
where 
\begin{eqnarray}
\varepsilon _\bot ^{\mu}=\varepsilon^{\mu \nu \rho \sigma}p_\nu  k_{1\rho }q_\sigma
\label{epst}
\end{eqnarray}
is the pseudovector with a normal direction to the scattering plane $({\bf k_1},{\bf k_2})$.
Our definitions of $\phi_h$ and other kinematic variables are in 
agreement with the common convention introduced in \cite{Bac2004}.

\begin{figure}[t]\centering
\vspace*{-6mm}
\scalebox{0.47}{\includegraphics{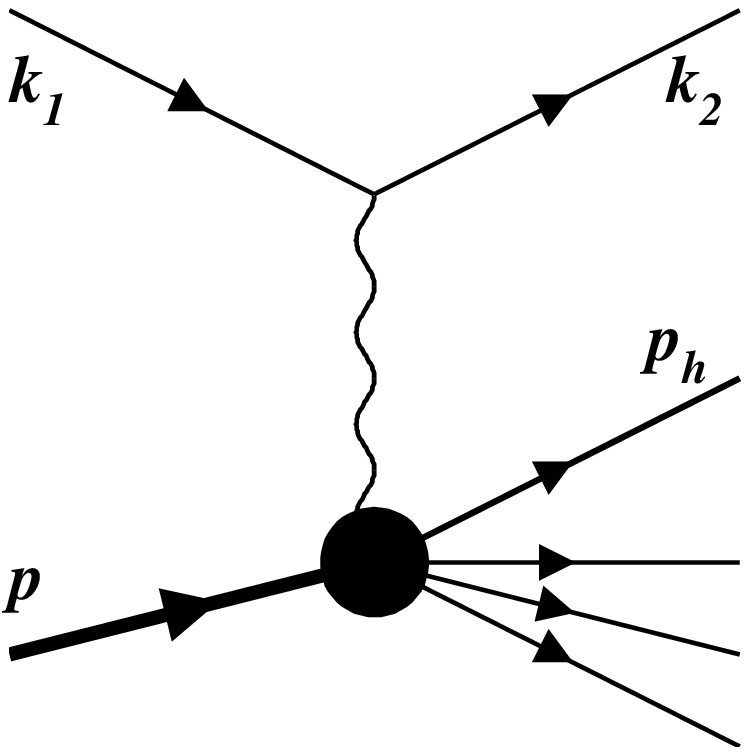}}
\\[-0.1cm]
{\bf a)}
\\[0.2cm]
\scalebox{0.47}{\includegraphics{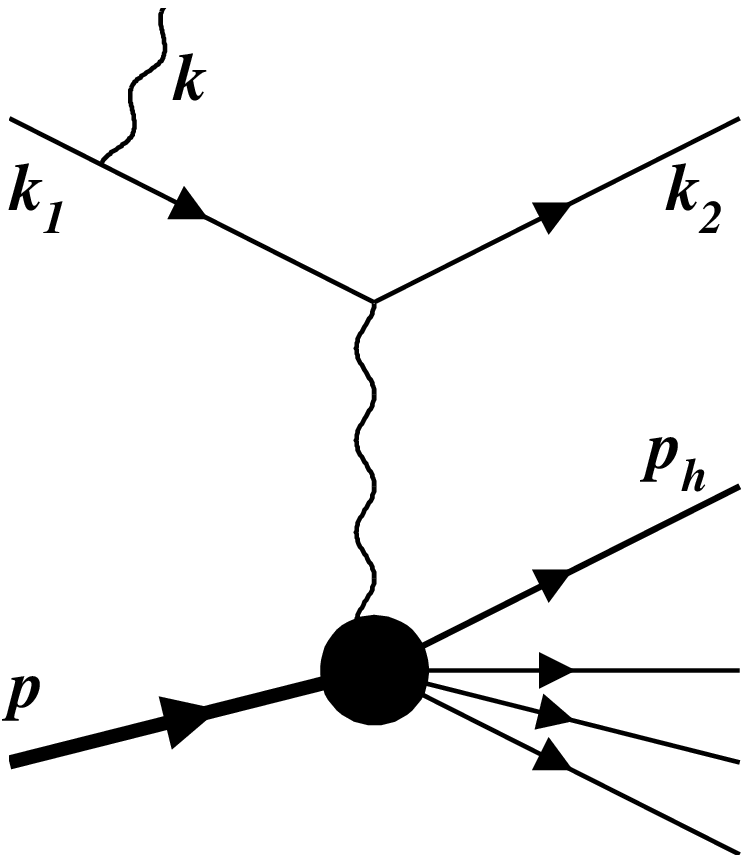}}
\scalebox{0.47}{\includegraphics{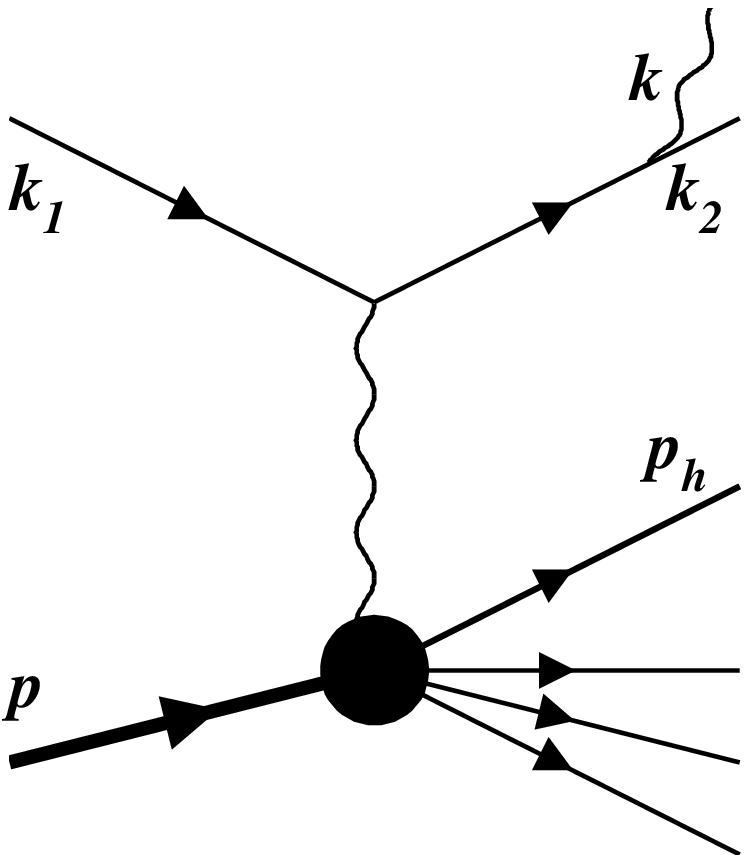}}
\\[-0.1cm]
{\bf \hspace{-.5cm} b) \hspace{3.3cm} c)\hspace{2.42cm}}
\\[0.1cm]
\scalebox{0.47}{\includegraphics{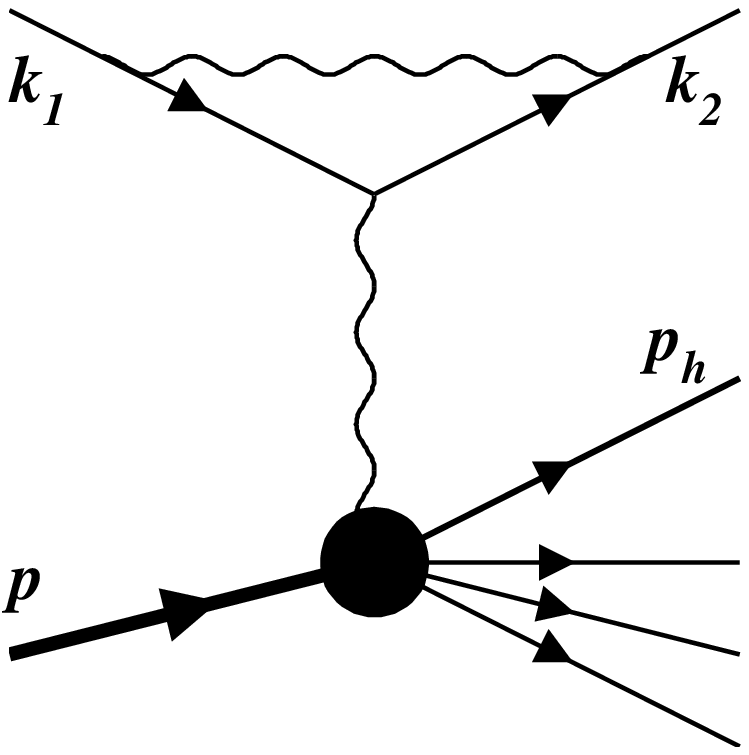}}
\scalebox{0.47}{\includegraphics{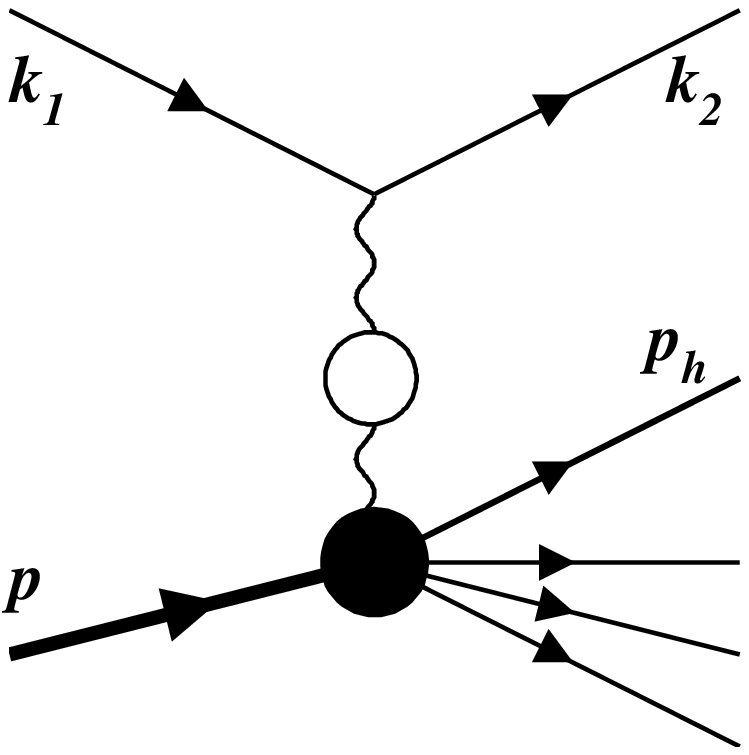}}
\\[-0.1cm]
{\bf \hspace{-.5cm} d) \hspace{3.3cm} e)\hspace{2.42cm}}
\\[0.1cm]
\scalebox{0.47}{\includegraphics{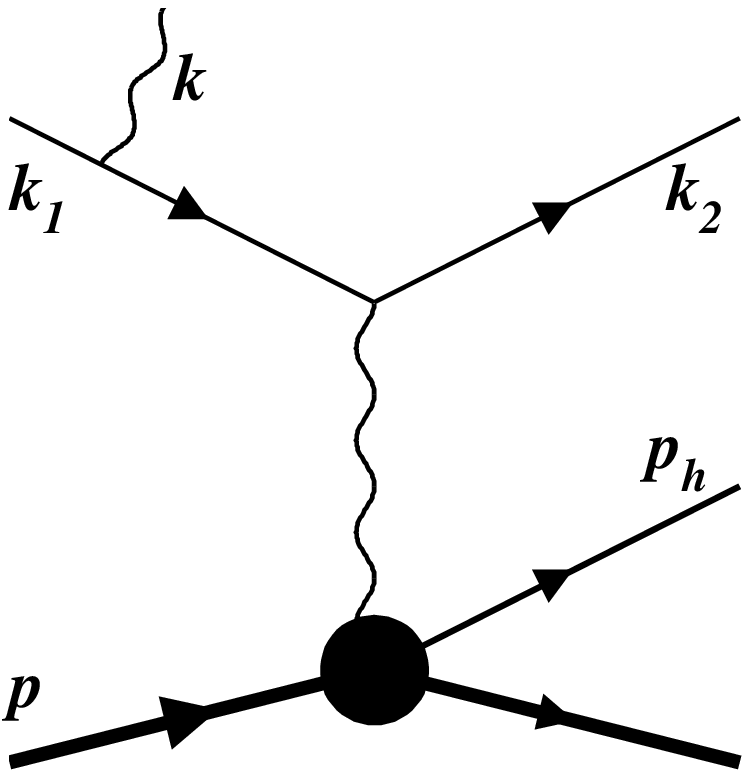}}
\scalebox{0.47}{\includegraphics{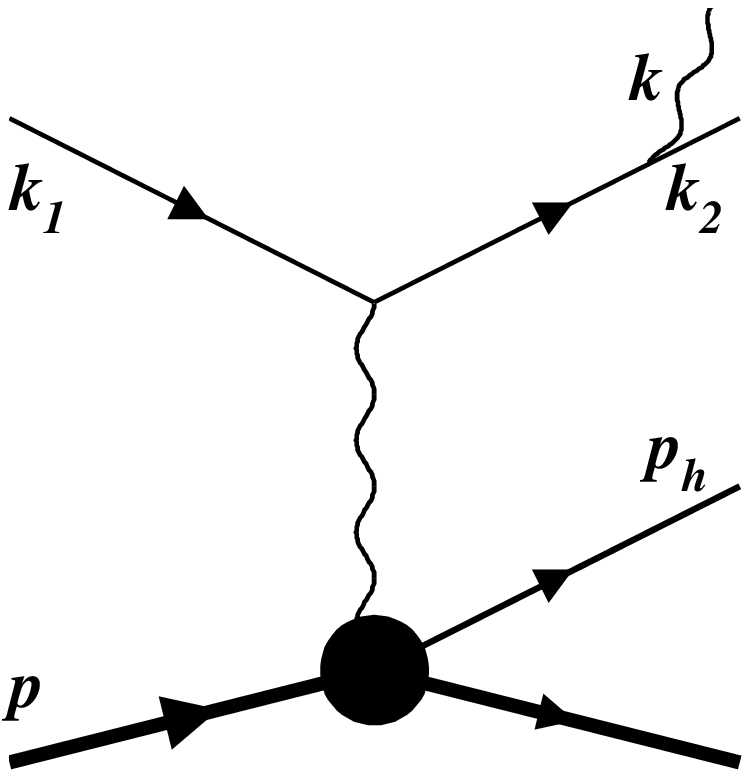}}
\\[-0.1cm]
{\bf \hspace{-.5cm} f) \hspace{3.3cm} g)\hspace{2.42cm}}
\\[0.1cm]
\caption{Feynman graphs for (a) the lowest order, (b)-(e) SIDIS and (f), (g)  exclusive  radiative tail 
contributions to the lowest order RC for SIDIS scattering} 
\label{fg}
\end{figure}

The lowest order QED (Born) contribution to SIDIS is presented by the Feynman graph in Fig.~\ref{fg} (a).
The cross section for this process reads
\begin{eqnarray}
d\sigma _B=\frac{(4\pi\alpha)^2}{2\sqrt{\lambda_S}Q^4}W_{\mu \nu}L^{\mu \nu}_B
d\Gamma _B,
\label{wl}
\end{eqnarray}
where the phase space is parametrized as
\begin{eqnarray}
d\Gamma _B&=&
(2\pi)^4\frac{d^3k_2}{(2\pi)^32k_{20}}
\frac{d^3p_h}{(2\pi)^32p_{h0}}
\nonumber\\
&=&\frac1{4(2\pi)^2}\frac {S S_xdx dy d\phi}{2\sqrt{\lambda_S}} \frac{S_xdzdp^2_td\phi_h}{4Mp_l}.
\end{eqnarray}

Since the initial lepton is considered to be longitudinally polarized, its
polarization vector has the form \cite{ASh}
\begin{eqnarray}
\xi  =\frac {\lambda_e S}{m\sqrt{\lambda_S}}k_{1}-\frac{2\lambda_e m}{\sqrt{\lambda_S}}p_1
=\xi _0+\xi_1 .
\label{xi}
\end{eqnarray}
As a result the leptonic tensor is
\begin{eqnarray}
L_B^{\mu \nu}&=&\frac 12 {\rm Tr}[({\hat k}_2+m)\gamma_{\mu }({\hat k}_1+m)(1+\gamma _5{\hat \xi})\gamma_{\nu}]
\nonumber\\
&=&2[k_{1}^{\mu}k_{2}^{\nu}+k_{2}^{\mu}k_{1}^{\nu}-\frac {Q^2}2g^{\mu \nu}
\nonumber\\&&
+\frac {i\lambda_e}{\sqrt{\lambda_S}} 
\varepsilon^{\mu \nu \rho \sigma}
(Sk_{2\rho} k_{1\sigma}+2m^2q_{\rho} p_{\sigma})
].
\label{lt0}
\end{eqnarray}

According to \cite{aram} the hadronic tensor for the SIDIS process $\gamma ^*+n\to h+X$ can be decomposed  in the terms of
the scalar spin-independent $H^{(0)}_{ab}$ and spin-dependent $H^{(S)}_{abi}$ structures functions
\begin{eqnarray}
W_{\mu \nu}=
\sum_{a,b=0}^3
e_\mu ^{\gamma (a)}
e_\nu ^{\gamma (b)}(H^{(0)}_{ab}+
\sum_{\rho,i=0}^3
\eta^\rho 
e^{h(i)}_\rho H^{(S)}_{abi}).
\label{ht0}
\end{eqnarray}
where $e_{\mu} ^{\gamma (a)}$ (or $e_{\nu} ^{\gamma (b)}$) and $e^{h(i)}_\rho$ 
are the complete set of the basis vectors for the polarization four-vectors of the virtual photon and nucleon 
in the target rest frame. These vectors can be represented in a covariant form \cite{arens} using (\ref{eps1}) and (\ref{e1}). 

Due to the parity and current conservation, hermiticity as well as $p\eta\equiv 0$, only the
following set of independent SF 
$H^{(0)}_{ab}$ and $H^{(S)}_{abi}$ in (\ref{ht0}) survives \cite{aram}:
5 spin-independent $H^{(0)}_{00}$, $H^{(0)}_{11}$, $H^{(0)}_{22}$, ${\rm Re}H^{(0)}_{01}$, ${\rm Im}H^{(0)}_{01}$
and 13 spin-dependent 
$H^{(S)}_{002}$, 
${\rm Re}H^{(S)}_{012}$, ${\rm Im}H^{(S)}_{012}$, ${\rm Re}H^{(S)}_{021}$, ${\rm Im}H^{(S)}_{021}$,
${\rm Re}H^{(S)}_{023}$, ${\rm Im}H^{(S)}_{023}$, $H^{(S)}_{112}$, 
${\rm Re}H^{(S)}_{121}$, ${\rm Im}H^{(S)}_{121}$,
${\rm Re}H^{(S)}_{123}$, ${\rm Im}H^{(S)}_{123}$, $H^{(S)}_{222}$. All the rest
of the SF have to be set to zero \cite{aram}.

The hadronic tensor in terms of these SF can be obtained by substitution (\ref{eps1}) and (\ref{e1}) into (\ref{ht0}) resulting in
\begin{eqnarray}
W_{\mu\nu}&=&\sum\limits_{i=1}^9 w^i_{\mu\nu}{\cal H}_i=
-g^\bot_{\mu \nu} {\cal H}_1
+p^\bot_\mu p^\bot_\nu {\cal H}_2
+p^\bot_{h\mu} p^\bot_{h\nu} {\cal H}_3
\nonumber\\&&
+(p^\bot_{\mu} p^\bot_{h\nu}+p^\bot_{h\mu} p^\bot_{\nu}) {\cal H}_4
+i(p^\bot_{\mu} p^\bot_{h\nu}-p^\bot_{h\mu} p^\bot_{\nu}) {\cal H}_5
\nonumber\\&&
+(p^\bot_{\mu} n_{\nu}+n_{\mu} p^\bot_{\nu}) {\cal H}_6
+i(p^\bot_{\mu} n_{\nu}-n_{\mu} p^\bot_{\nu}) {\cal H}_7
\nonumber\\&&
+(p^\bot_{h\mu} n_{\nu}+n_{\mu} p^\bot_{h\nu}) {\cal H}_8
\nonumber\\&&
+i(p^\bot_{h\mu} n_{\nu}-n_{\mu} p^\bot_{h\nu}) {\cal H}_9.
\label{ht1}
\end{eqnarray}
Here $g^\bot_{\mu \nu}=g_{\mu \nu}-q_{\mu }q_{\nu }/q^2$ and 
$n^{\mu}=\varepsilon^{\mu \nu \rho \sigma}q_\nu p_\rho p_{h\sigma}$.

The generalized SF ${\cal H}_i$ can be expressed via $H^{(0)}_{ab}$ and $H^{(S)}_{abi}$ 
  using the
decomposition of 
the nucleon polarized three-vector ${\boldsymbol \eta}=(\eta_1,\eta_2,\eta_3)$ 
over the basis (\ref{e1})
in the following way:
\begin{eqnarray}
{\cal H}_1&=&H^{(0)}_{22}-\eta_2H^{(S)}_{222}, 
\nonumber\\
{\cal H}_2&=&\frac 4{\lambda_Y^2 p_t^2}[\lambda_Y p_t^2Q^2(H^{(0)}_{00}-\eta _2H^{(S)}_{002})
+\lambda_3 ^2S_x^2(H^{(0)}_{11}
\nonumber\\&&
\qquad\;\;\;
-\eta_2H^{(S)}_{112})
-\lambda_2\lambda_Y (H^{(0)}_{22}-\eta_2H^{(S)}_{222}) 
\nonumber\\[2mm]&&
\qquad\;\;\;
-2 S_x \lambda_3p_tQ\sqrt{\lambda_Y}( {\rm Re} H^{(0)}_{01}-\eta_2 {\rm Re} H^{(S)}_{012})],
\nonumber\\[2mm]
{\cal H}_3&=&\frac 1{p_t^2}(H^{(0)}_{11}-H^{(0)}_{22}+\eta_2(H^{(S)}_{222}-H^{(S)}_{112})),
\nonumber\\
{\cal H}_4&=&\frac 2{\lambda_Y p_t^2}[\lambda_ 3 S_x(
H^{(0)}_{22}-H^{(0)}_{11}
+\eta_2(H^{(0)}_{112}-H^{(S)}_{222}))
\nonumber\\&&
\qquad\;\;\;
+p_tQ\sqrt{\lambda_Y}( {\rm Re}H^{(0)}_{01}-\eta_2 {\rm Re}H^{(S)}_{012})],
\nonumber\\
{\cal H}_5&=&\frac{2Q}{p_t\sqrt{\lambda_Y}}({\rm Im}H^{(0)}_{01}-\eta_2 {\rm Im}H^{(S)}_{012}),
\nonumber\\[2mm]
{\cal H}_6&=&\frac{4M}{\lambda_Y^{3/2}p_t^2}[
Qp_t\sqrt{\lambda_Y}(\eta_1{\rm Re}H^{(S)}_{021}+\eta_3{\rm Re}H^{(S)}_{023})
\nonumber\\&&
\qquad\;\;\;
-
\lambda_3S_x(\eta_1{\rm Re}H^{(S)}_{121}+\eta_3{\rm Re}H^{(S)}_{123})
],
\nonumber\\
{\cal H}_7&=&\frac{4M}{\lambda_Y^{3/2}p_t^2}[
Qp_t\sqrt{\lambda_Y}(\eta_1{\rm Im}H^{(S)}_{021}+\eta_3{\rm Im}H^{(S)}_{023})
\nonumber\\&&
\qquad\;\;\;
-
\lambda_3S_x(\eta_1{\rm Im}H^{(S)}_{121}+\eta_3{\rm Im}H^{(S)}_{123})
],
\nonumber\\
{\cal H}_8&=&\frac{2M}{\sqrt{\lambda_Y}p_t^2}(\eta_1{\rm Re}H^{(S)}_{121}+\eta_3{\rm Re}H^{(S)}_{123}),
\nonumber\\
{\cal H}_9&=&\frac{2M}{\sqrt{\lambda_Y}p_t^2}(\eta_1{\rm Im}H^{(S)}_{121}+\eta_3{\rm Im}H^{(S)}_{123}).
\end{eqnarray}
Here $\lambda_2=V_-^2+m_h^2Q^2$, $\lambda_3=V_-+zQ^2$, and $V_-$ is defined in Eqs.(\ref{setvar2}).

Finally we  find the Born contribution in the form
\begin{eqnarray}
\sigma^B\equiv \frac{d\sigma^B}{dxdydzdp_t^2d\phi_hd\phi}=
\frac{\alpha^2 SS^2_x}{8MQ^4p_l\lambda_S}\sum\limits_{i=1}^9\theta^B_i{\cal H}_i,
\label{born}
\end{eqnarray}
where $\theta^B_i=L^{\mu \nu}w^i_{\mu \nu}/2$,
\begin{eqnarray}
\theta^B_1&=&Q^2-2m^2,
\nonumber\\[1mm]
\theta^B_2&=&(SX-M^2Q^2)/2,
\nonumber\\[1mm]
\theta^B_3&=&(V_1V_2-m_h^2Q^2)/2,
\nonumber\\[1mm]
\theta^B_4&=&(SV_2+XV_1-zQ^2S_x)/2,
\nonumber\\[1mm]
\theta^B_5&=&\frac{2\lambda_eS\ea}{\sqrt{\lambda_S}} ,
\nonumber\\[1mm]
\theta^B_6&=&-S_p\ea ,
\nonumber\\[1mm]
\theta^B_7&=&
\frac {\lambda_eS}{4\sqrt{\lambda_S}}[\lambda_YV_+-S_pS_x(zQ^2+V_-)],
\nonumber\\[1mm]
\theta^B_8&=&-2V_+\ea ,
\nonumber\\[1mm]
\theta^B_9&=&\frac {\lambda_e}{2\sqrt{\lambda_S}}
[S (Q^2 (zS_x V_+-m_h^2S_p) + V_- (S V_2
\nonumber\\[1mm]&&
-XV_1))+2m^2(4M^2V_-^2+\lambda_Y m_h^2
\nonumber\\[1mm]&&
-z S_x^2 (z Q^2 +2 V_-))].
\label{thb}
\end{eqnarray}
\vspace*{-2mm}

The quantities $H^{(0)}_{ab}$ and $H^{(S)}_{abi}$
can be expressed through
another set of the SF presented in \cite{bacchetta}.
Taking into account that $\eta_1=\cos(\phi_s-\phi_h)S_\bot$, $\eta_2=\sin(\phi_s-\phi_h)S_\bot$
and  $\eta_3=S_{||}$ we find that
\begin{eqnarray}
&&
H^{(0)}_{00}=C_1F_{UU,L},\;
\nonumber \\&&
H^{(0)}_{01}=-C_1(F_{UU}^{\cos \phi_h}+iF_{LU}^{\sin \phi_h}),\;
\nonumber \\&&
H^{(0)}_{11}=C_1(F_{UU}^{\cos 2\phi_h}+F_{UU,T}),\;
\nonumber \\&&
H^{(0)}_{22}=C_1(F_{UU,T}-F_{UU}^{\cos 2\phi_h}),\;
\nonumber \\&&
H^{(S)}_{002}=C_1F_{UT,L}^{\sin (\phi_h-\phi_s) },\;
\nonumber \\&&
H^{(S)}_{012}=C_1(F_{UT}^{\sin \phi_s }-F_{UT}^{\sin (2\phi_h-\phi_s) }
\nonumber \\&&\qquad\qquad
-i(F_{LT}^{\cos \phi_s }-F_{LT}^{\cos (2\phi_h-\phi_s) })),\;
\nonumber \\&&
H^{(S)}_{021}=C_1(F_{UT}^{\sin (2\phi_h-\phi_s) }+F_{UT}^{\sin \phi_s }
\nonumber \\&&\qquad\qquad
-i(F_{LT}^{\cos (2\phi_h-\phi_s) }+F_{LT}^{\cos \phi_s })),\;
\nonumber \\&&
H^{(S)}_{023}=C_1(F_{UL}^{\sin \phi_h}-iF_{LL}^{\cos \phi_h}),\;
\nonumber \\&&
H^{(S)}_{121}=C_1(-F_{UT}^{\sin (3\phi_h-\phi_s) }-F_{UT}^{\sin (\phi_h+\phi_s) }
\nonumber \\&&\qquad\qquad
+iF_{LT}^{\cos (\phi_h-\phi_s) }),\;
\nonumber \\&&
H^{(S)}_{123}=C_1(-F_{UL}^{\sin 2\phi_h }+iF_{LL}),\;
\nonumber \\&&
H^{(S)}_{112}=C_1(F_{UT}^{\sin (3\phi_h-\phi_s) }+F_{UT,T}^{\sin (\phi_h-\phi_s) }
\nonumber \\&&\qquad\qquad
-F_{UT}^{\sin (\phi_h+\phi_s) }),\;
\nonumber \\&&
H^{(S)}_{222}=C_1(
F_{UT}^{\sin (\phi_h+\phi_s) }+F_{UT,T}^{\sin (\phi_h-\phi_s) }
\nonumber \\&&\qquad\qquad
-F_{UT}^{\sin (3\phi_h-\phi_s) }),\;
\end{eqnarray}
where
\begin{eqnarray}
C_1=\frac {4Mp_l(Q^2+2xM^2)}{Q^4}.
\end{eqnarray}

\section{\label{LO}Lowest Order Radiative Corrections}
The six matrix elements shown in Figs.~\ref{fg}(b)-\ref{fg}(g) contribute to the lowest order QED RC to the cross section of the base SIDIS process [Fig.~\ref{fg}(a)]. A critical difference in the graphs \ref{fg}(a)-\ref{fg}(e) comparing to the graphs \ref{fg}(f) and \ref{fg}(g) is the distinct final unobserved hadronic state: continuum of particles in the former case 
and a  single hadron in the latter case. The underlying processes are  semi-inclusive and exclusive hadron
leptoproduction, respectively. At the level of RC, both of them include the unobservable
real photon emission from the lepton leg as presented in 
Figs.~\ref{fg}(b) and \ref{fg}(c) as well as \ref{fg}(f) and \ref{fg}(g).
The contribution to RC from the semi-inclusive process contains also the
leptonic vertex correction
and 
vacuum polarization [Figs.~\ref{fg}(d) and \ref{fg}(e)]. Thus these two separate contributions to the total RC to the SIDIS cross section are considered in two separate subsections below.

\subsection{Semi-inclusive contribution}

The real photon emission in the semi-inclusive process, 
\begin{eqnarray}
e(k_1,\xi)+n(p, \eta)\to e(k_2)+h(p_h)+x(\tilde p_x)+\gamma(k),
\nonumber\\
\label{sir0}
\end{eqnarray}
where $k$ is a real photon four-momentum depicted in Figs.~\ref{fg}(b) and \ref{fg}(c)
is described by the set variables presented in~(\ref{setvar}) and three  additional quantities,
\begin{eqnarray}
R=2 kp,\; \tau=\frac{kq}{kp}, \; \phi_k,
\label{kvar}
\end{eqnarray}
where  $\phi_k$ is an angle between $({\bf k}_1,{\bf k}_2)$ and $({\bf k},{\bf q})$ planes. 
Its sine in the covariant form is
\begin{eqnarray}
\sin\phi_k=\frac{2\varepsilon _\bot k \sqrt{\lambda_Y}}
{R\sqrt{\lambda_1(Q^2+\tau (S_x-\tau M^2))}}. 
\label{phik}
\end{eqnarray}

The contribution of real photon emission from the leptonic leg is
\begin{eqnarray}
d\sigma_R&=&\frac{(4\pi\alpha)^3}{2\sqrt{\lambda_S}{\tilde Q}^4}{\widetilde W}_{\mu \nu}L_{R}^{\mu \nu}
d\Gamma_R.
\label{sr}
\end{eqnarray}
Here the ``tilde'' symbol denotes that the arguments of the hadronic tensor such as $Q^2$, $W^2$, $z$, $t$ and $\phi_h$   
are defined
through the shifted $q \to q-k$, {\it i.e.} 
${\tilde Q}^2=-(q-k)^2=Q^2+R\tau$. The phase space of the considered process has the form  
\begin{eqnarray}
d\Gamma_R=(2\pi)^4\frac{d^3k}{(2\pi)^32k_{0}}
\frac{d^3k_2}{(2\pi)^32k_{20}}
\frac{d^3p_h}{(2\pi)^32p_{h0}},
\end{eqnarray}
where
\begin{eqnarray}
\frac{d^3k}{k_{0}}=\frac{RdRd\tau d\phi _k}{2\sqrt{\lambda _Y}}.
\end{eqnarray}

For the representation of explicit results in the simplest way
the leptonic tensor $L_{R}^{\mu \nu}$ in (\ref{sr}) is separated into two parts:
\begin{eqnarray}
L_{R}^{\mu \nu}=L_{R0}^{\mu \nu}+L_{R1}^{\mu \nu}. 
\label{ltr}
\end{eqnarray}
The first term includes the part of the leptonic tensor that contains spin-independent terms and terms containing $\xi_0$, i.e., the part of the polarization vector (\ref{xi}),
\begin{eqnarray}
L_{R0}^{\mu \nu }&=&-\frac 12 {\rm Tr}[({\hat k}_2+m)\Gamma^{\mu \alpha}_R
({\hat k}_1+m)(1+\gamma _5{\hat \xi}_0){\bar \Gamma}^{\nu}_{R\alpha}],
\nonumber\\
\label{lr0}
\end{eqnarray}
where
\begin{eqnarray}
\Gamma^{\mu \alpha }_R&=&\Biggl(\frac{k_1^{\alpha}}{kk_1}-\frac{k_2^{\alpha}}{kk_2} \Biggr)\gamma ^\mu
-\frac{\gamma ^\mu{\hat k}\gamma ^{\alpha }}{2kk_1}-\frac{\gamma ^\alpha {\hat k}\gamma ^{\mu }}{2kk_2},
\nonumber\\
{\bar \Gamma^{\nu}_{R\alpha }}&=&\gamma_0\Gamma^{\nu \dagger}_{R \alpha }\gamma_0
\nonumber\\&=&
\Biggl(\frac{k_{1\alpha}}{kk_1}-\frac{k_{2\alpha}}{kk_2} \Biggr)\gamma ^\nu
-\frac{\gamma ^\nu{\hat k}\gamma _{\alpha }}{2kk_2}-\frac{\gamma _\alpha {\hat k}\gamma ^{\nu }}{2kk_1}.
\end{eqnarray}
The second term in (\ref{ltr}) is proportional only to the residual part $\xi_1 $ of the polarization vector $\xi $,
\begin{eqnarray}
L_{R1}^{\mu \nu }&=&-\frac 12 {\rm Tr}[({\hat k}_2+m)\Gamma^{\mu \alpha}_R({\hat k}_1+m)\gamma _5{\hat \xi}_1 
{\bar \Gamma}^{\nu}_{R \alpha}].
\label{lrp}
\end{eqnarray}
As shown below this part of the leptonic tensor
gives a nonvanishing contribution to RC in the ultrarelativistic 
approximation for both the semi-inclusive (\ref{sin1}) and
the exclusive  (\ref{exc1}) final hadronic states.

The convolution of the leptonic tensors $L_{R0}^{\mu \nu}$ and $L_{R1}^{\mu \nu}$
with the shifted hadronic tensor can be presented as   
\begin{eqnarray}
{\widetilde W}_{\mu \nu}L_{R0}^{\mu \nu}&=&\sum\limits_{i=1}^{9}{\widetilde w}^i_{\mu \nu}{\tilde {\cal H}}_iL_{R0}^{\mu \nu}=
-2\sum\limits_{i=1}^{9}\sum\limits_{j=1}^{k_i}{\tilde {\cal H}}_i\theta^0_{ij}R^{j-3},
\nonumber\\
{\widetilde W}_{\mu \nu}L_{R1}^{\mu \nu}&=&
\sum\limits_{i=5,7,9}{\widetilde w}^i_{\mu \nu}{\tilde {\cal H}}_iL_{R1}^{\mu \nu}=
-2\sum\limits_{i=5,7,9}\sum\limits_{j=1}^{k_i}{\tilde {\cal H}}_i\theta^1 _{ij}R^{j-3},
\nonumber\\
\label{l0rw}
\end{eqnarray}
where $i$ enumerates the contributions of respective SF in (\ref{ht1}). The sum over $j$ represents the 
decomposition of the leptonic (\ref{lr0}) and (\ref{lrp}) and hadronic tensor convolutions over $R$. 
In this decomposition quantities $\theta^{0,1} _{ij}$ do not depend on $R$. Their explicit expressions 
are presented in Appendix~
\ref{ap1}. The number of terms is different for different SF: $k_i=\{3,3,3,3,3,4,4,4,4\}$.

The lowest order SIDIS process~(\ref{lo}) 
 is described by the four independent four-momenta such us $p$, $k_1$, $q$ and $p_h$. Therefore, the Born cross section contains only one pseudoscalar $\varepsilon^{\mu \nu \rho \sigma}p_{h\;\mu}p_\nu k_{1\rho}q_\sigma $. This pseudoscalar contributes to $\theta^B_{5,6,8}$ as it was shown in Eqs.~(\ref{thb}) and, according to Eqs.~(\ref{sphih}) and (\ref{epst}), can be expressed in terms of the variables (\ref{setvar})-(\ref{setvar3}) as 
$\varepsilon^{\mu \nu \rho \sigma}p_{h\;\mu}p_\nu k_{1\rho}q_\sigma =\varepsilon_\bot p_h 
=-p_t\sqrt{\lambda_1}\sin\phi_h /2$. When we deal with real photon emission, the additional independent four-momentum $k$ appears. 
As a result the number of
pseudoscalar quantities that can exist in the expressions for the cross section grows up to five. They are not independent and their number can be reduced to two,
namely $\varepsilon_\bot p_h$ and $\varepsilon_\bot k$, using the decomposition of 
the photonic four-momentum over the basis introduced in Appendix~\ref{ap0} by Eqs.~(\ref{e2}). As shown in Eqs.~(\ref{deceps}) the remaining three pseudoscalars are expressed through the linear
combination of $\varepsilon_\bot p_h$ and $\varepsilon_\bot k$. 
The explicit expression for $\varepsilon_\bot k$ follows from: 
(\ref{phik})
\begin{eqnarray}
\varepsilon _\bot k=\frac{\sin\phi_k R\sqrt{\lambda_1(Q^2+\tau (S_x-\tau M^2))}}{2\sqrt{\lambda_Y}}.
\label{phik2}
\end{eqnarray}

After substituting (\ref{l0rw}) into (\ref{sr}) 
\begin{eqnarray}
d\sigma_R&=&-\frac{\alpha^3}{4\pi^2{\tilde Q}^4 \sqrt{\lambda_S}}
\sum\limits_{i=1}^{9}\sum\limits_{j=1}^{k_i}{\tilde {\cal H}}_i\theta_{ij}R^{j-3}
\frac{d^3k}{k_{0}}
\frac{d^3k_2}{k_{20}}
\frac{d^3p_h}{p_{h0}}
\nonumber\\
&=&
-\frac{\alpha ^3S S_x^2
dxdydzdp_td\phi_hd\phi d\tau d\phi_k dR  }{64\pi ^2M p_l\lambda_S\sqrt{\lambda _Y}{\tilde Q}^4}
\nonumber\\&&\times
\sum\limits_{i=1}^9
\sum\limits_{j=1}^{k_i}{\tilde {\cal H}}_i\theta_{ij}
R^{j-2},
\label{sr2}
\end{eqnarray}
where 
$\theta _{ij}=\theta _{ij}^0$ for $i=1-4,6,8$ and
$\theta _{ij}=\theta _{ij}^0+\theta _{ij}^1$ for $i=5,7,9$,
we found that the term with $j=1$ in (\ref{sr2}) contains the infrared divergence at $R\to 0$
that does not allow one to perform the straightforward integration of $d\sigma_R$ over the photonic variable $R$. 
For the correct extraction and cancellation of the infrared divergence 
the Bardin-Shumeiko approach \cite{BSh} is used. 
Following this method the identical transformation,
\begin{eqnarray}
d\sigma_R=d\sigma_R-d\sigma_R^{IR}+d\sigma_R^{IR}=d\sigma_R^F+d\sigma_R^{IR},
\label{irff}
\end{eqnarray}
is performed.
Here $d\sigma_R^F$ is the infrared free contribution
and $d\sigma_R^{IR}$ contains only the $j=1$ term in which arguments of SF are taken for $k=0$,
\begin{eqnarray}
d\sigma_R^{IR}&=&-\frac{\alpha^3}{4\pi^2Q^4 \sqrt{\lambda_S}}
\sum\limits_{i=1}^{9}\frac{{\cal H}_i\theta_{i1}}{R^2}
\frac{d^3k}{k_{0}}
\frac{d^3k_2}{k_{20}}
\frac{d^3p_h}{p_{h0}}.
\label{ir}
\end{eqnarray}
This decomposition allows us to perform the treatment of the infrared divergence analytically
since the arguments of the SF in (\ref{ir}) do not depend on photonic
variables.
Due to $\theta_{i1}=4 F_{IR}\theta^B_i$
one can find that this contribution can be factorized in front of the Born cross section 
\begin{eqnarray}
d\sigma_R^{IR}
=-\frac{\alpha}{\pi^2}d\sigma^B\frac{F_{IR}}{R^2}\frac{d^3k}{k_{0}},
\label{sir}
\end{eqnarray}
where
\begin{eqnarray}
F_{IR}=\biggl(\frac {k_1}{z_1}-\frac {k_2}{z_2}\biggr )^2,
\end{eqnarray}
$z_{1,2}=kk_{1,2}/kp$, and the explicit expressions 
of these quantities are given in Appendix \ref{ap1} [see (\ref{z1z2})]. 

The term (\ref{sir}) is then separated
into the soft $\delta_S$ and hard $\delta_H$ parts,
\begin{eqnarray}
\sigma_R^{IR}=\frac {\alpha}{\pi}(\delta_S+\delta_H)\sigma ^B
\label{sir2}
\end{eqnarray}
by the introduction of the infinitesimal photonic energy 
${\bar k}_0\to 0$ that is defined in the system ${\bf p}+{\bf q}-{\bf p_h}=0$:
\begin{eqnarray}
\delta_S&=&-\frac 1\pi\int \frac{d^3k}{k_0}\frac{F_{IR}}{R^2}\theta ({\bar k}_0-k_0),
\nonumber\\[1mm]
\delta_H&=&-\frac 1\pi\int \frac{d^3k}{k_0}\frac{F_{IR}}{R^2}\theta (k_0-{\bar k}_0).
\label{dsh}
\end{eqnarray}

The additional regularization with the parameter $\bar k_0$ allows us to calculate $\delta_H$ for
$n=4$
and to simplify the integration for $\delta_S$ in the dimensional regularization by choosing the individual reference systems for each invariant variables $z_1$ and $z_2$
to make them independent of the azimuthal angle $\phi_k$.

The explicit integration, details of which are described in Appendix \ref{ap2}, results in  
the final explicit expressions for these two contributions in the form
\begin{eqnarray}
\delta_S&=&2(Q_m^2L_m-1)\biggl( P_{IR}+\log \frac {2{\bar k}_0}\nu \biggr)
+\frac 12S^\prime L_{S^\prime}
\nonumber\\[1mm]&&
+\frac 12X^\prime L_{X^\prime}
+ S_\phi,
\nonumber\\[1mm]
\delta_H&=&2(Q_m^2L_m-1)\log \frac{p_x^2-M_{th}^2}{2{\bar k}_0\sqrt{p_x^2}}.
\label{ird}
\end{eqnarray}
Here $M_{th}$ is the minimum value of the invariant mass of the undetected hadrons $p_x$ for the SIDIS process, e.g., $M_{th}=M+m_\pi$ when the detected hadron is the pion. The symbols $L_m$, $L_{S^\prime}$, and $L_{X^\prime}$ are defined in Eq.~(\ref{lms}).

The sum of $\delta_S$ and $\delta_H$ does not depend on
the separated photonic energy ${\bar k}_0$ but includes the term representing the infrared divergence
\begin{eqnarray}
P_{IR}=\frac 1{n-4}+\frac 12\gamma _E+\log\frac 1{2\sqrt{\pi}}
\label{pir}
\end{eqnarray}
as well as the arbitrary parameter $\nu $, the mass scale of dimensional regularization. These two quantities
should be canceled by summing the infrared divergent part 
with the contribution from the leptonic vertex correction
 that is
considered below.

The term $S_\phi$ has the form
\begin{eqnarray}
S_\phi&=&-\frac{Q_m^2}{2\sqrt{\lambda_m}}
\biggl \{\log\frac{X^\prime-\sqrt{\lambda_X^\prime}}{X^\prime+\sqrt{\lambda_X^\prime}}
\log\frac{(z-z_1)(z-z_3)}{(z-z_2)(z-z_4)}
\nonumber \\&&
+\sum_{i,j}^{4}S_j(-1)^{i+1}
\biggl(\frac 12\delta_{ij}\log ^2(|z-z_i|)
\nonumber \\&&
+(1-\delta_{ij})\biggl[\log(|z-z_i|)
\log (|z_i-z_j|)
\nonumber \\&&
-{\rm Li}_2\biggl(\frac {z-z_i}{z_j-z_i}\biggr)\biggr]\biggr)\biggr\}{\bigg\vert}_{z=z_d}^{z=z_u}, 
\label{sph0}
\end{eqnarray}
where 
\begin{eqnarray}
{\rm Li}_2(x)=-\int\limits^x_0\frac{\log|1-y|}y dy
\end{eqnarray}
is Spence's dilogarithm and
\begin{eqnarray}
z_{1,2}&=&\frac 1{\sqrt{\lambda_X^\prime}}
\biggl(X^\prime-S^\prime+\frac{2p_x^2(Q^2\mp\sqrt{\lambda_m})}{X^\prime-\sqrt{\lambda_X^\prime}}\biggr),
\nonumber \\
z_{3,4}&=&\frac 1{\sqrt{\lambda_X^\prime}}
\biggl(S^\prime-X^\prime-\frac{2p_x^2(Q^2\pm\sqrt{\lambda_m})}{X^\prime+\sqrt{\lambda_X^\prime}}\biggr),
\nonumber \\
z_u&=&\sqrt{\frac{\lambda_S^\prime}{\lambda_X^\prime}}-1,\;
z_d=\frac{X^\prime (S^\prime-X^\prime)-2p_x^2Q^2}{\lambda_X^\prime},
\nonumber \\
S_j&=&\{1,1,-1,-1\}.
\end{eqnarray}

The infrared free contribution $d\sigma^F_R$ from (\ref{irff}) integrated over the three photonic variables 
reads
\begin{eqnarray}
\sigma_R^F&=&-\frac{\alpha ^3S S_x^2}{64\pi ^2M p_l\lambda_S\sqrt{\lambda _Y}
}
\int\limits_{\tau _{\rm min}}^{\tau _{\rm max}}
d\tau  
\int\limits_{0}^{2\pi}
d\phi_k  
\int\limits_{0}^{R _{\rm max}}
dR  
\nonumber\\[1mm]&&\times
\sum\limits_{i=1}^9
\Biggl[\frac{\theta_{i1}}R
\Biggl(\frac {{\tilde {\cal H}}_i}{{\tilde Q}^4}-\frac {{\cal H}_i}{Q^4}\Biggr)+
\sum\limits_{j=2}^{k_i}{\tilde {\cal H}}_i\theta_{ij}\frac{R^{j-2}}{{\tilde Q}^4}\Biggr],
\nonumber\\
\label{srfin}
\end{eqnarray}
where the limits of integration are
\begin{eqnarray}
R_{\rm max}&=&\frac {p_x^2-M_{th}^2}{1+\tau-\mu },
\nonumber\\
\tau _{\rm max/min}&=&\frac{S_x\pm \sqrt{\lambda_Y}}{2M^2}
\end{eqnarray}
and the quantity $\mu$ is defined in Eq.~(\ref{mudef}).

The additional virtual particle contributions consist of the leptonic vertex correction [Fig.\ref{fg} (d)]
and vacuum polarization by leptons and hadrons [Fig.\ref{fg} (e)].
These contributions are given by Eq.~(\ref{wl})
with the replacement of the leptonic tensor  $L_B^{\mu \nu}$ by
\begin{eqnarray}
L_{V}^{\mu \nu}&=&
\frac 12 {\rm Tr}[({\hat k}_2+m)\Gamma_{V}^{\mu }({\hat k}_1+m)(1+\gamma _5{\hat \xi})\gamma^{\nu}]
\nonumber \\&&
+
\frac 12 {\rm Tr}[({\hat k}_2+m)\gamma^{\mu }({\hat k}_1+m)(1+\gamma _5{\hat \xi}){\bar \Gamma}^{\nu}_{V} ],
\end{eqnarray}
where 
\begin{eqnarray}
\Gamma_{V}^{\mu }=\Lambda^{\mu }+
\Pi^{l\mu}_\alpha 
\gamma^\alpha 
+
\frac \alpha {2\pi } \delta_{\rm vac}^h
\gamma^\mu
, 
\end{eqnarray}
and ${\bar \Gamma}^{\nu}_{V}=\gamma_0\Gamma^{\nu\;\dagger}_{V}\gamma_0$.

The first two terms corresponding to the leptonic vertex correction  
$\Lambda_{\mu }$ and vacuum polarization by leptons $\Pi^{l\mu} _\alpha $ 
are calculated analytically using Feynman rules while the fit for the vacuum polarization 
by hadrons $\delta_{\rm vac}^h$ can be taken from 
the experimental data \cite{vach}.

Since $\Lambda_{\mu }$ and $\Pi^{l\mu} _\alpha $
contain the ultraviolet divergence while
$\Lambda_{\mu }$ also includes the infrared divergent 
term the dimensional regularization is used for the calculation of the loop integrals:
\begin{eqnarray}
\Lambda_{\mu }&=&-ie^2\int \frac{d^nl}{(2\pi)^n\nu^{n-4}}
\nonumber \\&&\times
\frac{\gamma_\alpha ({\hat k}_2-{\hat l}+m)\gamma_\mu({\hat k}_1-{\hat l}+m)\gamma^\alpha}{l^2(l^2-2lk_2)(l^2-2lk_1)}, 
\nonumber \\
\Pi^l_{\alpha \mu }&=&-\frac{ie^2}{Q^2}\int \frac{d^nl}{(2\pi)^n\nu^{n-4}}
\nonumber \\&&\times\Biggl\{ \sum_{i=e,\mu,\tau}
\frac{{\rm Tr} [({\hat l}+m_i)\gamma_\alpha({\hat l}-{\hat q}+m_i)\gamma_\mu]}{(l^2-m_i^2)((l-q)^2-m_i^2)}
\Biggr \}.
\nonumber\\
\label{lp0}
\end{eqnarray}
Details of the calculations are presented in Appendix~\ref{ap3};
$\Lambda_{\mu }$ and $\Pi^i_{\alpha \mu} $ have the following structure:
\begin{eqnarray}
\Lambda_{\mu }&=&\frac {\alpha}{2\pi}
\biggl(\delta_{\rm vert}^{UV}(Q^2)\gamma _\mu-\frac 12 mL_m[{\hat q},\gamma _\mu]\biggr),  
\nonumber \\
\Pi^l_{\alpha \mu }&=&\sum_{i=e,\mu,\tau}\frac {\alpha}{2\pi}
\delta_{\rm vac}^{i\; UV}(Q^2)g^\bot_{\alpha \mu},
\label{lp2}
\end{eqnarray}
where the second term in $\Lambda_{\mu }$ is the anomalous magnetic moment. 
To remove the ultraviolet divergence 
the standard on the mass-shell renormalization procedure is used:  $\delta_{\rm vert}^{UV}(Q^2)$ and $\delta_{\rm vac}^{i\; UV}(Q^2)$ are substituted by the difference of these quantities and their values at $Q^2=0$:
\begin{eqnarray}
\delta_{\rm vert}=\delta_{\rm vert}^{UV}(Q^2)-\delta_{\rm vert}^{UV}(0),
\nonumber\\
\delta_{\rm vac}^i=\delta_{\rm vac}^{i\; UV}(Q^2)-\delta_{\rm vac}^{i\; UV}(0).  
\end{eqnarray}
Here $\delta_{\rm vert}^{UV}(0)=2-P_{UV}-2P_{IR}-3\log (m/\nu)$, $\delta_{\rm vac}^{i\; UV}(0)=4(P_{UV}+\log (m_i/\nu))/3$, $P_{UV}=P_{IR}$, 
and the ultraviolet free terms have the form
\begin{eqnarray}
\label{dvrtvac}
\delta_{\rm vert}&=&-2(Q^2_mL_m-1)\biggl(P_{IR}+\log\frac m\nu \biggr)-2  
\nonumber \\&&
+\biggl( \frac 32 Q^2+4m^2 \biggr)L_m-\frac{Q_m^2}{\sqrt{\lambda_m}}
\biggl(\frac 12\lambda_mL_m^2
\nonumber \\&&
+2{\rm Li}_2\biggl(\frac{2\sqrt{\lambda_m}}{Q^2+\sqrt{\lambda_m}} \biggr)-\frac{\pi^2}2 \biggr), 
\nonumber \\
\delta_{\rm vac}^l&=&\sum_{i=e,\mu,\tau}\delta_{\rm vac}^i=\sum_{i=e,\mu,\tau}
\Bigl[\frac 23(Q^2+2m^2_i)L_m^i
\nonumber\\&&
-\frac {10} 9
+\frac{8m_i^2}{3Q^2}\Bigl(1-2m^2_iL_m^i
\Bigr)
\Bigr].
\end{eqnarray}
The quantity 
 $L_m$ is defined in (\ref{lms}) while the expressions for $\lambda_m^i$ and $L_m^i$ are defined by Eqs.~(\ref{lmi}).

Finally the contribution of the inelastic tail to the sixfold SIDIS cross section
reads
\begin{eqnarray}
\sigma^{in}&=&
\frac{\alpha }{\pi }(\delta_{VR}
+\delta_{\rm vac}^l
+\delta_{\rm vac}^h
)\sigma^{B}
+\sigma^F_R
+\sigma^{\rm AMM},
\label{srv}
\end{eqnarray}
where the sum of the infrared divergent terms,
\begin{eqnarray}
\delta_{VR}&=&\delta_S+\delta_H+\delta_{\rm vert}
\nonumber \\
&=&2(Q_m^2L_m-1)\log \frac{p_x^2-M_{th}^2}{m \sqrt{p_x^2}}
+\frac 12S^\prime L_{S^\prime}
\nonumber \\&&
+\frac 12 X^\prime L_{X^\prime}+S_{\phi }-2  
+\biggl( \frac 32 Q^2+4m^2 \biggr)L_m
\nonumber \\&&
-\frac{Q_m^2}{\sqrt{\lambda_m}}
\biggl(\frac 12\lambda_mL_m^2
+2{\rm Li}_2\biggl(\frac{2\sqrt{\lambda_m}}{Q^2+\sqrt{\lambda_m}} \biggr)
\nonumber \\&&
-\frac{\pi^2}2 \biggr) 
\label{dvr}
\end{eqnarray}
is free both from the infrared divergent term $P_{IR}$ appearing in $\delta_S$ and $\delta_{\rm vert}$
that are defined by Eqs.~(\ref{ird}) and (\ref{dvrtvac}) and the arbitrary parameter $\nu$.
The infrared free contribution $\sigma^F_R$ is defined by Eq.~(\ref{srfin}).

At last the contribution of the anomalous magnetic moment
coming from the second term in $\Lambda_{\mu }$ given by Eqs.~(\ref{lp2})
has the form
\begin{eqnarray}
\sigma^{\rm AMM}=\frac{\alpha^3m^2 SS^2_x}{16\pi MQ^2p_l\lambda_S}
L_m\sum\limits_{i=1}^9\theta^{\rm AMM}_i{\cal H}_i,
\label{amm}
\end{eqnarray}
with
\begin{eqnarray}
\theta^{\rm AMM}_1&=&6,
\nonumber \\
\theta^{\rm AMM}_2&=&-\frac{\lambda_Y} {2Q^2},
\nonumber \\
\theta^{\rm AMM}_3&=&-2m_h^2-2\frac{V_-^2} {Q^2},
\nonumber \\
\theta^{\rm AMM}_4&=&-2S_x\biggl( z+\frac{V_-} {Q^2}\biggr),
\nonumber \\
\theta^{\rm AMM}_5&=&\frac{2\lambda_e(2S+S_x)\ea } {\sqrt{\lambda_S}Q^2},
\nonumber \\
\theta^{\rm AMM}_7&=&\frac{\lambda_e(2S+S_x)} {4\sqrt{\lambda_S}Q^2}
(S_x(SV_2-XV_1-zS_pQ^2)
\nonumber \\&&
+4M^2Q^2V_+),
\nonumber \\
\theta^{\rm AMM}_9&=&\frac{\lambda_e} {2\sqrt{\lambda_S}Q^2}
(S_x^2(4m^2(m_h^2-z(zQ^2+2V_-))
\nonumber \\&&
+V_1V_-)-4(M^2(Q^2-4m^2)+S^2)(m_h^2Q^2
\nonumber \\&&
+V_-^2)
+zQ^2S_x(S_x(zQ^2+V_1+V_-)
\nonumber \\&&
+2SV_+)
+2SS_xV_- V_+),
\nonumber \\
\theta^{\rm AMM}_6&=&\theta^{\rm AMM}_8=0.
\end{eqnarray}

\subsection{Exclusive radiative tail}
The exclusive radiative tail is the process
 \begin{eqnarray}
e(k_1,\xi)+n(p, \eta)\to e(k_2)+h(p_h)+u(p_u)+\gamma(k),
\nonumber\\
\label{excr}
\end{eqnarray}
where $p_u$ is the four-momentum of a single undetected hadron ($p_u^2=m_u^2$) 
shown in Figs.~\ref{fg}(f) and \ref{fg} (g). 
The final unobserved state contains the photon radiated from the lepton line and a hadron produced in an exclusive reaction of $\gamma^*$ and $p$.
The process (\ref{excr}) gives a contribution to the RC in SIDIS because two observed particles in the final state can have the same momenta as the unobserved particles in the
SIDIS process~(\ref{lo}). The square of the invariant mass of the unobserved state $p_x^2=(p+q-p_h)^2=2k(p+q-p_h)+m_u^2$ depends on the photonic variables. Emission of the soft photons would result in $p_x^2=m_u^2$. This is beyond the kinematic region of SIDIS. Therefore 
the process (\ref{excr}) being the contribution to RC to the SIDIS cross section does not contain the infrared divergence 
\cite{AIO}.

A description of the exclusive process without the radiated photon requires  
only five of the six variables of SIDIS presented in Eqs.~(\ref{setvar}): $x$, $y$, $t$, $\phi_h$, and $\phi$. 
The process with the radiated photon is additionally described by the three photonic variables
$R$, $\tau $, and $\phi_k$ introduced above by Eq.~(\ref{kvar}). In this case the sixth SIDIS variable $z$ is
expressed through other SIDIS and photonic variables:
\begin{eqnarray}
z=\frac{M^2-m_u^2+t-R(1+\tau-\mu)}{S_x}+1,
\end{eqnarray}
where $\mu$ is defined by Eq.~(\ref{mudef}). Since we calculate RC to SIDIS we need to keep $z$ and use this equation in order to express $R$ in terms of $z$ and two remaining photonic variables:
\begin{eqnarray}
R_{ex}=\frac{p_x^2-m_u^2}{1+\tau-\mu},
\end{eqnarray}
and therefore to reduce the integration over the photon momentum to the two-dimensional integral with respect to
variables $\tau $ and $\phi_k$.

 The contribution of the exclusive radiative tail in the form similar to (\ref{sr}) reads
\begin{eqnarray}
d\sigma^{ex}_R=\frac{(4\pi \alpha) ^3}{2\sqrt{\lambda _S}\tilde Q^4}\widetilde W^{\mu \nu}_{ex}L^R_{\mu \nu}d\Gamma _R^{ex},
\label{sre}
\end{eqnarray}
where the hadronic tensor $W^{\mu \nu}_{ex}$ describes the exclusive process $\gamma ^*+n\to h+u$
and has the same structure as the hadronic tensor in Eq.~(\ref{ht1}) 
but with the SF dependent only on $Q^2$, $W^2$ and $t$
variables. 
The leptonic tensor $L^{R}_{\mu \nu}$
and its convolution with the hadronic structures $\tilde w_i^{\mu \nu}$ are the same as in Eqs.~(\ref{lr0})-(\ref{l0rw}). 

The phase space of this process is
\begin{eqnarray}
d\Gamma _R^{ex}&=&\frac 1{(2\pi)^8}\frac{d^3k_2}{2k_{20}}\frac{d^3k}{2k_{0}}\frac{d^3p_h}{2p_{h0}}\frac{d^3p_u}{2p_{u0}}
\nonumber\\&&\times
\delta ^4(k_1+p-k_2-p_h-p_u-k)
\nonumber\\
&=&\frac{2R_{ex} SS_x^2dxdyd\phi dz d\phi_h dp_t^2d\tau d\phi_k}{(4\pi)^8(1+\tau -\mu )Mp_l\sqrt{\lambda_S\lambda_Y}}. 
\label{dgf}
\end{eqnarray}

The use of  the phase space (\ref{dgf}) and convolution of leptonic and hadronic tensors (\ref{l0rw})
with replacement
${\tilde \mathcal H}_i \to {\tilde \mathcal H}^{ex}_i$   in  (\ref{sre}) and the subsequent integration of
the obtained expression over two photonic variables
results in the contribution of the exclusive radiative tail to the SIDIS process in the form
\begin{eqnarray}
\sigma^{ex}_{R}&=&-\frac{\alpha^3SS_x^2}{2^9\pi^5Mp_l\lambda_S\sqrt{\lambda_Y}}
\int\limits_{\tau _{\rm min}}^{\tau _{\rm max}}d\tau
\int\limits_0^{2\pi}d\phi_k
\nonumber \\&&\times
\sum_{i=1}^{9}
\sum_{j=1}^{k_i}
\frac{{\tilde \mathcal H}^{ex}_i\theta_{ij}R_{ex}^{j-2}}{(1+\tau-\mu)\tilde Q^4}.
\label{sre1}
\end{eqnarray}

\section{\label{URA} Ultrarelativistic Approximation}
In  Sec. \ref{LO} all contributions to the lowest order RC are presented by exact formulas. 
Some of them have a rather complicated analytical structure.
However, due to the smallness of the leptonic mass  compared to other
quantities that describe kinematics of the process it is rather useful to obtain RC in 
the ultrarelativistic approximation keeping the leptonic mass $m$ only as an argument of the logarithmic function.
This allows us to simplify the analytical expressions essentially as well as clarify the leading log 
behavior of the obtained
results. In other words, the lowest order QED RC in this approximation has the form
\begin{eqnarray}
\sigma_{RC}&=&\frac{\alpha }{\pi}\Biggl[A l_m+B+{\mathcal O}\Biggl(\frac{m^2}{Q^2}\Biggr)\Biggr],
\label{urarc}
\end{eqnarray}
where $l_m=\log Q^2/m^2 $ and the terms $A$ and $B$ are independent of the leptonic mass and represent the lowest order leading and next-to-leading
contributions to the RC to the cross section, respectively. 

The terms in (\ref{srv}) that are 
 factorized in front of the Born contribution are essentially simplified, resulting in a more transparent structure after applying the ultrarelativistic approximation, e.g., the terms (\ref{sir2}) 
\begin{eqnarray}
\sigma_R^{IR}&=&\frac {\alpha}{\pi}\biggl[
(l_m-1)\biggl(2P_{IR}+2\log\frac m\nu
\nonumber\\&&
+\log\frac{(p_x^2-M_{th}^2)^2}{S^\prime X^\prime}
\biggr)+\frac 12 l_m^2
\nonumber\\&&
-\frac 12 \log ^2 \frac {S^\prime }{X^\prime }
+{\rm Li}_2
\biggl\{
1-\frac{Q^2p_x^2}{S^\prime X^\prime}
\biggr\}-\frac {\pi ^2}3
\biggr]\sigma_0
\end{eqnarray}
contain both $l_m$ and $l_m^2$. The latter comes from the soft photon emission whose contribution
cancels in the sum with the leptonic vertex correction: 
\begin{eqnarray}
\delta_{VR}&=&(l_m-1)\log\frac{(p_x^2-M_{th}^2)^2}{S^\prime X^\prime}
+\frac 32l_m
\nonumber\\&&
-\frac 12 \log ^2 \frac {S^\prime }{X^\prime }
+{\rm Li}_2
\biggl\{
1-\frac{Q^2p_x^2}{S^\prime X^\prime}
\biggr\}-\frac {\pi ^2}6-2.
\label{dvru}
\end{eqnarray}

The vacuum polarization by lepton $i$ ($i=e,\mu,\tau $) in the limit $Q^2\gg m_i^2$ reads
\begin{eqnarray}
\delta_{\rm vac}^i&=&\frac 23\log\frac {Q^2}{m^2_i}-\frac {10} 9.
\label{pvu}
\end{eqnarray}

The ultrarelativistic approximation for
 the hard photon emission contribution (\ref{srfin},\ref{sre1}) requires additional care because of the integration over photonic variables and the nontrivial dependence of the integrand on the leptonic mass. Specifically, the integrand contains the terms $1/z_1$ and $1/z_1^2$:
\begin{eqnarray}
&\displaystyle
\int \limits _0^{2\pi } \frac{d\phi_k}{z_1}=
\frac{2\pi \sqrt{\lambda _Y }}{\sqrt{(Q^2+\tau S)^2+4m^2(\tau (S_x-\tau M^2)+Q^2)}},
\nonumber\\[1mm]
&\displaystyle
\int \limits _0^{2\pi } \frac{d\phi_k}{z_1^2}=
\frac{2\pi (Q^2S_p+\tau (S S_x+2M^2Q^2))\sqrt{\lambda _Y }}{((Q^2+\tau S)^2+4m^2(\tau (S_x-\tau M^2)+Q^2))^{3/2}}
\nonumber\\
\label{int1}
\end{eqnarray}
These have a sharp peaking behavior in the region
$\tau \to \tau_s\equiv-Q^2/S$ due to the smallness of the lepton mass. The integration of the expressions (\ref{int1}) over $\phi_k$ and $\tau $ gives \
\begin{eqnarray}
&\displaystyle
\int \limits_{\tau _{\rm min}}^{\tau _{\rm max}}d\tau \int \limits _0^{2\pi } \frac{d\phi_k}{z_1}
=2\pi\sqrt{\frac{\lambda _Y }{\lambda _S }}\log\frac{S+\sqrt{\lambda _S }}{S-\sqrt{\lambda _S }},
\nonumber\\[1mm]
&\displaystyle
\int \limits_{\tau _{\rm min}}^{\tau _{\rm max}}d\tau \int \limits _0^{2\pi } \frac{d\phi_k}{z_1^2}=
\frac{2\pi\sqrt{\lambda _Y }}{m^2}.
\label{int2}
\end{eqnarray}

Since
\begin{eqnarray}
\lim_{m\to 0}\log\frac{S+\sqrt{\lambda _S }}{S-\sqrt{\lambda _S }}=l_m+\log\frac{S^2}{Q^2M^2}
\end{eqnarray}
the terms containing $1/z_1$ contribute to the leading and next-to-leading RC. The terms containing $1/z_1^2$ also contain $m^2$ in numerators and therefore contribute to the next-to-leading RC only (the only exception is $\hat\theta^0_{53}$ that is discussed below). The similar conclusions
are true for the terms containing  $1/z_2$ and $1/z_2^2$ terms. 

Actually the integrand in (\ref{int2}) has to be multiplied by the
SF according to (\ref{srfin}) and (\ref{sre1}). Therefore, we make the identical transformation for extraction of the leading and next-to-leading terms:
\begin{eqnarray}
&\displaystyle
\int \limits_{\tau _{\rm min}}^{\tau _{\rm max}}d\tau \int \limits _0^{2\pi } d\phi_k 
\frac{{\cal G}(\tau,\phi_k)}{z_1}
=
2\pi\sqrt{\frac{\lambda _Y }{\lambda _S }}\log\frac{\sqrt{\lambda _S }+S}{\sqrt{\lambda _S }-S}
{\cal G}(\tau _s,0)
\nonumber\\[1mm]
&\displaystyle
+
\int \limits_{\tau _{\rm min}}^{\tau _{\rm max}}d\tau \int \limits _0^{2\pi } d\phi_k 
\frac{{\cal G}(\tau,\phi_k)-{\cal G}(\tau _s,0)}{z_1},
\nonumber\\
&\displaystyle
m^2\int \limits_{\tau _{\rm min}}^{\tau _{\rm max}}d\tau \int \limits _0^{2\pi } d\phi_k 
\frac{{\cal G}(\tau,\phi_k)}{z_1^2}
=
2\pi\sqrt{\lambda _Y }
{\cal G}(\tau _s,0)
\nonumber\\[1mm]
&\displaystyle
+
\int \limits_{\tau _{\rm min}}^{\tau _{\rm max}}d\tau \int \limits _0^{2\pi } d\phi_k 
m^2\frac{{\cal G}(\tau,\phi_k)-{\cal G}(\tau _s,0)}{z_1^2},
\label{int21}
\end{eqnarray}
where ${\cal G}(\tau,\phi_k)$
is a regular function of $\tau$ and $\phi_k$. 
The second term in the right-hand side of the first transformation does not include the leading terms and the second term in the
second equality is proportional to $m^2$ and vanishes in the ultrarelativistic
approximation. 

The approach of extraction of the leading and next-to-leading contributions can be illustrated by considering the terms originated from the convolution of the leptonic tensor (\ref{lrp}) with the hadronic structures ${\widetilde w}^i_{\mu \nu}$.
Summing up the terms $\theta^1_{ij} R^{j-3} $ in the last expression of Eq.~(\ref{l0rw})  
and keeping the leptonic  mass only in the term $m^2/z_1^2$ (in $\theta^1_{ij}$ the term $1/z_2^2$ is proportional to $m^4$)
results in
\begin{eqnarray}
{\widetilde w}^i_{\mu \nu}L_{ R1}^{\mu \nu}=-2\sum\limits_{j=1}^{k_i}\theta_{ij}^1R^{j-3}
= \frac {m^2}{z_1^2}\theta_i^1(R,\tau,\phi_k)
\label{lprw}
\end{eqnarray}
with the quantities $\theta_i^1(R,\tau_s,0)$ expressed
through (\ref{thb}) as
\begin{eqnarray}
\theta_i^1(R,\tau_s,0)=\frac{4R}{S(S-R)}\theta_i^B\Biggl(k_1\rightarrow \biggl(1-\frac RS\biggr)k_1\Biggr).
\end{eqnarray}
The replacement in the brackets is applied for any kinematic variable defined through $k_1$, e.g., $S\to S-R$, $Q^2\to (1-S/R)Q^2$, and $\ea \to (1-S/R)\ea$. 
 Note that $R=R_{ex}$ has to be used for the exclusive radiative tail.

The resulting equation for the $\sigma_R^{\xi_1}$ is obtained using the second equation of (\ref{int21}) with the regular function  ${\cal G}(\tau,\phi_k)$,
\begin{eqnarray}
{\cal G}(\tau,\phi_k)=\int\limits_0^{R_{\rm max}}\frac{RdR}{(Q^2+\tau R)^2}
\sum_{i=5,7,9}\theta_i^1(R,\tau,\phi_k)
{\tilde {\cal H}}_i.
\nonumber \\
\end{eqnarray}
Therefore, the contribution from the second part $\xi_1$ of the lepton polarized vector (\ref{xi})
reads
\begin{eqnarray}
\sigma_R^{\xi_1}=-\frac {\alpha S_x^2}{\pi M p_lS^2}\int\limits_{0}^{R^s_{\rm max}}\frac {p_l^sRdR}{(S_x-R)^2}{\tilde \sigma} ^B_{pl},
\label{sin1}
\end{eqnarray}
where 
\begin{eqnarray}
p_l^s&=&\frac{z SS_x(S_x-R)+2M^2(RV_1-2SV_-)}{2M\sqrt{S(4M^2Q^2(S-R)+S(S_x-R)^2)}},
\nonumber\\
R^s_{\rm max}&=&S(p_x^2-M_{th}^2)/S^\prime ,
\end{eqnarray}
and ${\tilde \sigma} ^B_{pl}$ is proportional to the $\lambda_e$ part of the Born contribution with the following replacement:
$m\to0$, $S\to S-R$, $Q^2\to Q^2(1-R/S)$, $V_1\to V_1(1-R/S)$  and $z\to z S_x/(S_x-R)$.

A similar calculation of the exclusive radiative tail results in
\begin{eqnarray}
\sigma_R^{ex\;\xi_1}
=-\frac {\alpha S_x^2R^s_{ex}p_l^{s\;ex}}{\pi M p_lSS^{\prime }(S_x-R^s_{ex})}
\frac{d{\tilde \sigma} ^{ex\;B}_{pl}}{d\tilde xd\tilde yd\tilde p_td\phi_hd\phi},
\label{exc1}
\end{eqnarray}
where 
\begin{eqnarray}
p_l^{s\; ex}&=&
\frac 1{2M\sqrt{S(4M^2Q^2(S-R^s_{ex})+S(S_x-R^s_{ex})^2)}}
\nonumber\\&&
[(S_x-R^s_{ex})(S(S_x-2V_-+m_h^2-m_u^2)
\nonumber\\&&
-R^s_{ex}(S-V_1))
-Q^2(S-R^s_{ex})(S_x-R^s_{ex})
\nonumber\\&&
+M^2(
S(S_x-4V_-)-
R^s_{ex}(S-2V_1)
)
],
\end{eqnarray}
$R^s_{ex}=S(p_x^2-m_u^2)/S^{\prime}$,
and the exclusive Born cross section reads
\begin{eqnarray}
\frac{d\sigma ^{ex\;B}_{pl}}{dxdydp_td\phi_hd\phi}=\frac{\alpha^2 S S_x}{64\pi^3Q^4M p_l\lambda_S}
\qquad
\qquad
\qquad
\nonumber\\\times
\sum_{i=5,7,9}{\mathcal H}^{ex}_i\theta^B_i(z\to \frac{t+M^2-m_u^2}{S_x}+1).
\end{eqnarray}

Finally, we consider the extraction of the leading and next-to-leading terms in the quantity ${\hat \theta}^0_{53}$ given in Appendix~\ref{ap1}. 
In contrast to other ${\hat \theta}^0_{ij}$, the quantity ${\hat \theta}^0_{53}$ includes terms $1/z_1^2$ without factors proportional to $m^2$ and therefore can potentially result in 
electron mass singularity $\sim m^{-2}$ after integration (\ref{int2}). This is, however, not the case because ${\hat \theta}_{53}^0=0$ at the peak point, i. e., for $\tau=\tau _s=-Q^2/S$ (and $\mu=V_1/S$). Explicit integration in the limit $m^2\to 0$, 
\begin{widetext}
\begin{eqnarray}
\int\limits_{\tau_{\rm min}}^{\tau_{\rm max}}d\tau\int\limits_0^{2\pi}d\phi_k {\hat \theta}_{53}^0
=-\frac{2\lambda_e \pi p_t\sin\phi_h\sqrt{\lambda_Y}}{M^2S^2\sqrt{Q^2(S X-M^2Q^2)}}
\biggl[ 4M^2Q^2(S X-M^2Q^2)\biggl(l_m+\log\frac {S^2}{Q^2M^2}-3\biggr)+S^2\lambda_Y  \biggr]
\end{eqnarray}
\end{widetext}
shows that ${\hat \theta}^0_{53}$ has a standard form $A\log(Q^2/m^2)+B$.

The final result for the observed cross section in the ultrarelativistic approximation
is obtained by the following substitutions in Eqs.(\ref{srv}) and (\ref{sre1}): 
(i) $\sigma^{\rm AMM}=0$; 
(ii) Eqs.~(\ref{dvru}) and (\ref{pvu}) for $\delta_{VR}$ and $\delta_{\rm vac}^i$; 
(iii) $m=0$ in the Born cross section [Eq. (\ref{born})]; and 
(iv) 
$\sigma_R^{F}=\sigma_R^{F1}+\sigma_R^{\xi_1}$ and 
$\sigma_R^{ex}=\sigma_R^{ex\;1}+\sigma_R^{ex \;\xi_1}$, 
where $\sigma_R^{\xi_1}$ and $\sigma_R^{ex\;\xi_1}$ are given by (\ref{sin1}) and (\ref{exc1}),
respectively, and $\sigma_R^{F1}$ and $\sigma_R^{ex\; 1}$ are given by Eqs. (\ref{srfin}) and (\ref{sre1}) with $\theta^1_{ij}=0$ and the leptonic mass keeping only in the coefficients at $F_{21}$ and $F_{22}$.

\section{\label{Conc} Conclusion}

Newly achieved accuracies in modern SIDIS experiments in TJNAF and CERN require 
renewed attention to RC calculations and their implementation in data analysis software.  In this paper we obtained the exact analytical expressions for the lowest order model-independent part of
QED RC to the SIDIS
cross section with the longitudinally polarized initial lepton and arbitrarily polarized target and demonstrated how the leading and next-to-leading contributions can be extracted.
The model-independent RC includes (i) the contributions of radiated SIDIS processes and loop diagrams (\ref{srv})
 and (ii) the contribution of the exclusive radiative tail (\ref{sre1}). The methodology developed in this paper is the extension of the covariant approach for the RC calculations developed earlier: (i) the method of covariant extraction and cancellation of the infrared divergence suggested by Bardin and Shumeiko \cite{BSh}; (ii) the set of integration variables used in RC calculation to DIS \cite{ASh}; (iii) RC to unpolarized and polarized SIDIS in the quark-parton model \cite{Polrad,SSh1,SSh2}; (iv) RC for SIDIS of unpolarized particles \cite{ASSh}; and (v) the calculation of the exclusive radiative tail for unpolarized SIDIS \cite{AIO}. The calculations of RC in SIDIS measurements were performed by the model-independent way that involves constructing and using the SIDIS (and exclusive) hadronic tensor containing the 18 SIDIS and exclusive SF. We obtained the explicit form of the hadronic tensor using the approaches of \cite{aram} and \cite{arens} and demonstrated that the Born cross section exactly coincides with that given by \cite{bacchetta}. The next step in the RC calculation includes coding of the formulas and numeric evaluation of the effects of the RC. However, this requires models of the SIDIS/exclusive SF that are not known now. Therefore, a broad discussion and efforts of theoreticians and experimentalists are required to complete the evaluation of all SIDIS SF as well as SF in the resonance region and exclusive SF. Further development will include development of (i) iteration procedure with fitting of measured SF and joining with models beyond SIDIS measurements at each iteration step, and (ii) tools for generation of the radiated photon. Such a generator can be constructed based on a code for RC in SIDIS in the same way of how RADGEN \cite{RADGEN} is constructed based on POLRAD 2.0. Generation of semi-inclusive processes based on DIS Monte Carlo generators can provide only approximate cross sections, because a part of the SIDIS cross section involving pure semi-inclusive SF and respective convolutions of the leptonic and hadronic tensors are not presented in such DIS Monte Carlo generators.

\acknowledgments
The authors are grateful to Harut Avakian and Andrei Afanasev for interesting discussions and comments. This
work was supported by DOE Contract No. DE-AC05-06OR23177, under which Jefferson Science Associates,
LLC, operates Jefferson Lab.

\appendix
\section{Bases in the four-dimensional space}
\label{ap0}
 In this appendix three bases in the four-dimensional space that are used in 
 our analyses, are presented. The first two are used for the decomposition of the initial target and virtual photon polarization in the hadronic tensor defined by (\ref{ht0}). The latter allows us to decompose the real photon momentum
in such a way that all five pseudoscalar quantities appearing in processes (\ref{sir0}) and (\ref{excr}) reduce down to two: $\ea$ 
and $\varepsilon_\bot k$. 

For the decomposition of the hadronic tensor over the SF
it is convenient to introduce the reference system $({\bf x}_h,{\bf y}_h,{\bf z}_h)$ 
in the target rest frame where the two polar axes are defined as follows: ${\bf z}_h$ is chosen in the virtual 
 photon three-momentum direction
${\bf q}={\bf k}_1-{\bf k}_2$
and the ${\bf x}_h$ along the part of the registrated hadronic momentum that is transverse to the ${\bf z}_h$ axis.
The direction of the rest axial ${\bf y}_h$ axis is defined as ${\bf y}_h={\bf z}_h\times {\bf x}_h$.
In this system
the complete basis for polarization vectors can be presented in covariant form \cite{arens}
for both the virtual photon 
\begin{eqnarray}
e ^{\gamma (0)}_\mu&=&\frac{2Q}{\sqrt{\lambda_Y}}p^\bot_\mu,
\nonumber \\[1mm]
e ^{\gamma (1)}_\mu&=&\frac 1{p_t}\Biggl[
p^\bot_{h \mu}
-
\frac {S_x(m_h^2+(2z-1)Q^2-t)}{\lambda_Y}p^\bot_\mu
\Biggr],
\nonumber \\[1mm]
e ^{\gamma (2)}_\mu&=&
2\frac{\varepsilon^{\mu \nu \rho \sigma}p_\nu q_\rho p_{h\sigma}}{p_t\sqrt{\lambda_Y}},
\nonumber \\
e ^{\gamma (3)}_\mu&=&\frac{q_\mu}{Q},
\label{eps1}
\end{eqnarray}
and the
nucleon
\begin{eqnarray}
e^{h(0)}_\mu&=&\frac {p_\mu} M,
\nonumber \\[1mm]
e^{h(1)}_\mu&=&\frac 1{p_t}\Biggl[
p^\bot_{h \mu}
-
\frac {S_x(m_h^2+(2z-1)Q^2-t)}{\lambda_Y}p^\bot_\mu
\Biggr],
\nonumber \\[1mm]
e^{h(2)}_\mu&=&
2\frac{\varepsilon^{\mu \nu \rho \sigma}p_\nu q_\rho p_{h\sigma}}{p_t\sqrt{\lambda_Y}},
\nonumber \\
e^{h(3)}_\mu&=&\frac {2M^2 q_\mu-S_xp_\mu}{M\sqrt{\lambda_Y}},
\label{e1}
\end{eqnarray}
where $Q=\sqrt{Q^2}$ as well as for any four-vector $a^\bot_\mu=a_\mu+aq \; q_\mu/Q^2$.
Note that the direction of $e^{h(2)}$ (and $e^{\gamma (2)}$ as well)  is chosen in such a way that
the projection of ${\bf k}_{1,2}$ on ${\bf y}_h$ reads 
${\bf y}_h\cdot {\bf k}_1={\bf y}_h\cdot {\bf k}_2=-e^{h(2)}k_1=-e^{h(2)}k_2=k_t\sin(\phi_h )$. 

The components of these two bases in the reference system $({\bf x}_h,{\bf y}_h,{\bf z}_h)$ read
\begin{eqnarray}
e ^{\gamma (0)}_\mu&=&\frac1{2MQ}(\sqrt{\lambda_Y},0,0,S_x),
\;e^{h(0)}_\mu=(1,0,0,0),
\nonumber \\[1mm]
e^{\gamma (1)}_\mu&=&(0,1,0,0),
\qquad \qquad \;\;\;\;\; 
e^{h(1)}_\mu=(0,1,0,0),
\nonumber \\[1mm]
e^{\gamma (2)}_\mu&=&(0,0,1,0),
\qquad \qquad \;\;\;\;\; 
e^{h(2)}_\mu=(0,0,1,0),
\nonumber \\
e^{\gamma (3)}_\mu&=&\frac1{2MQ}(S_x,0,0,\sqrt{\lambda_Y}),
\;e^{h(3)}_\mu=(0,0,0,1).
\label{eps12}
\end{eqnarray}

In the rest frame system the virtual photon
longitudinal and transverse polarizations correspond
to $e ^{\gamma (0)}$ and $e ^{\gamma (1,2)}$, respectively,
and the left and right circular polarizations
are defined as
\begin{eqnarray}
e ^{\gamma (\pm)}=\mp\frac 1{\sqrt{2}}(e ^{\gamma (1)}\pm ie ^{\gamma (2)}).
\end{eqnarray}

To decompose the photonic four-momentum the other reference system $({\bf x}_l,{\bf y}_l,{\bf z}_l)$ 
in the rest target frame has to be introduced. In this system the polar ${\bf z}_l$ axis
has the same direction as the three-vector ${\bf q}$, the other polar ${\bf x}_l$ axis is chosen along 
the incoming or outgoing
lepton part that is
transverse to ${\bf q}$, and the axial ${\bf y}_l$ axis is defined as ${\bf y}_l={\bf z}_l\times {\bf x}_l$.
As a result  $({\bf x}_l,{\bf y}_l)$  is the scattering plane. In the covariant form this basis reads as  
\begin{eqnarray}
e^{l(0)}_\mu&=&\frac {p_\mu} M,
\nonumber \\[1mm]
e^{l(1)}_\mu&=&\sqrt{\frac {\lambda_Y }{\lambda_1}}\Biggl[
\frac 12(k_{1\mu}+k_{2 \mu})-\frac {S_pQ^2}{\lambda_Y}p^{\bot }_\mu\Biggr],
\nonumber \\[1mm]
e^{l(2)}_\mu&=&
-\frac{2\varepsilon_{\bot \mu}}{\sqrt{\lambda_1}},
\nonumber \\
e^{l(3)}_\mu&=&\frac {2M^2 q_\mu-S_xp_\mu}{M\sqrt{\lambda_Y}}.
\label{e2}
\end{eqnarray}
Note that the direction of ${\bf y}_l$  is chosen in such a way that
the projection of ${\bf p}_h$ on ${\bf y}_l$ is ${\bf y}_l\cdot {\bf p}_h=-e^{l(2)}p_h=-p_t\sin(\phi_h )$. 
The two reference systems $({\bf x}_{h},{\bf y}_{h},{\bf z}_{h})$ and
 $({\bf x}_{l},{\bf y}_{l},{\bf z}_{l})$ can be expressed through 
each other in the following way: 
\begin{eqnarray}
{\bf x}_{h}&=&{\bf x}_{l} \cos (\phi_h)-{\bf y}_{l}\sin (\phi_h),
\nonumber \\
{\bf y}_{h}&=&{\bf x}_{l} \sin (\phi_h)+{\bf y}_{l} \cos (\phi_h),
\nonumber \\
{\bf z}_{h}&=&{\bf z}_{l}
\end{eqnarray}
where $\cos (\phi_h)$ and $\sin (\phi_h)$ are defined by Eqs.~(\ref{cphih}) and (\ref{sphih})
respectively.

It should also be noted that for $i=\gamma, h,l$
\begin{eqnarray}
e^{i (a)}_\mu e ^{i (b)}_\nu g^{\mu \nu}=g^{a b},
\nonumber \\
e^{i (a)}_\mu e ^{i (b)}_\nu g_{a b}=g_{\mu \nu}.
\end{eqnarray}

The photonic four-momentum can be decomposed into the following way, $k=k_{(a)}e^{(a)}$, where
\begin{eqnarray}
k_{(0)}&=&ke^{l(0)}=\frac R{2M}, 
\nonumber\\
\nonumber\\
k_{(1)}&=&-ke^{l(1)}=\frac {R(Q^2S_p+\tau (S S_x+2M^2Q^2)-z_1\lambda_Y)}{2\sqrt{\lambda_1\lambda_Y}}, 
\nonumber \\[1mm]
k_{(2)}&=&-ke^{l(2)}=\frac {2\varepsilon _\bot k }{\sqrt{\lambda_1}}, 
\nonumber \\[1mm]
k_{(3)}&=&-ke^{l(3)}=\frac {R (S_x-2\tau M)}{2M\sqrt{\lambda_1}}. 
\end{eqnarray}

This decomposition for the four-momentum of the real unobservable photon  allows us to express all pseudoscalars
through the linear combinations of two of them, $\ea$ and $\varepsilon _\bot k$:
\begin{eqnarray}
\varepsilon^{\mu \nu \rho \sigma}k_\mu p_{h \;\nu}  k_{1\rho }q_\sigma&=&
\frac 1 {2\lambda_1}(R \varepsilon _\bot p_h
(\tau (Q^2S+2m^2S_x)
\nonumber \\[1mm]
&&+Q^2(4m^2+Q^2-z_1S_p))
\nonumber \\[1mm]
&&+\varepsilon _\bot k(Q^2(S V_2+X V_1-z Q^2 S_x)
\nonumber \\[1mm]
&&-4m^2S_x(z Q^2+V_-))),
\nonumber \\[1mm]
\varepsilon^{\mu \nu \rho \sigma}k_\mu p_\nu  p_{h \;\rho }q_\sigma&=&
\frac 1 {2\lambda_1}(R \varepsilon _\bot p_h(z_1\lambda_Y
-Q^2S_p
\nonumber \\[1mm]&&
-\tau (S S_x+2M^2Q^2)
)
\nonumber \\[1mm]&&
+\varepsilon _\bot k
(S_x(z Q^2 S_p-S V_2+X V_1)
\nonumber \\[1mm]&&
-4V_+M^2Q^2)),
\nonumber \\[1mm]
\varepsilon^{\mu \nu \rho \sigma}k_\mu p_\nu k_{1\rho }  p_{h \;\sigma}&=&
\frac 1 {2\lambda_1}(R \varepsilon _\bot p_h
(\tau \lambda_S+2m^2S_x
\nonumber \\[1mm]&&
+Q^2S-z_1(SS_x+2M^2Q^2))
\nonumber \\[1mm]&&
+\varepsilon _\bot k
(2m^2(4V_-M^2-z S_x^2)
\nonumber \\[1mm]&&
+S(S V_2-X V_1-z Q^2 S_x)
\nonumber \\[1mm]&&
+2V_1M^2Q^2)).
\label{deceps}
\end{eqnarray}

\section{Explicit expression for $\theta _{ij}$}
\label{ap1}
For all $i=1-8$, the quantities $\theta^0_{i1}=4 F_{IR}\theta^B_i$ and $F_{IR}$ are defined by (\ref{fff}). The other $\theta^0_{ij}$ read
\begin{widetext}
\begin{eqnarray}
\theta^0 _{12}&=&4\tau F_{IR},
\nonumber\\[1mm]
\theta^0 _{13}&=&-4-2 F_{d}\tau ^2,
\nonumber\\[1mm]
2\theta^0 _{22}&=&S_xS_pF_{1+}+2m^2S_pF_{2-}+2(S_x-2\tau M^2 )F_{IR}-\tau S_p^2F_d, 
\nonumber\\[1mm]
2\theta^0 _{23}&=&(4m^2+\tau (2\tau M^2-S_x))F_d-S_pF_{1+}+4M^2,
\nonumber\\[1mm]
\theta^0 _{32}&=&2((\mu V_--\tau m_h^2)F_{IR}+V_+(\mu m^2F_{2-} +V_-F_{1+}-\tau V_+F_d)),
\nonumber\\[1mm]
\theta^0 _{33}&=&(2\mu ^2m^2+ \tau (\tau m_h^2-\mu V_-))F_d-\mu V_+F_{1+}+2m_h^2,
\nonumber\\[1mm]
\theta^0 _{42}&=&(SV_1-X V_2)F_{1+}+m^2(\mu S_p+2V_+)F_{2-}-2\tau S_pV_+F_d
+((\mu-2 \tau z)S_x+2V_-)F_{IR},
\nonumber\\[1mm]
2\theta^0 _{43}&=&(8\mu m^2+ \tau ((2\tau z-\mu)S_x-2V_-))F_d-(\mu S_p+2V_+)F_{1+}
+4zS_x, 
\nonumber\\[1mm]
\theta^0 _{52}&=&
\frac {\lambda_ eS}{\lambda_ 1\sqrt{\lambda_ S}}
[\ea 
(2(S_x(Q^2+4m^2)+2\tau (S X- 2M^2 (Q^2+2 m^2)))F_{IR}
+Q^2(S_p (S_xF_{1+}+2m^2 F_{2-})
\nonumber\\[1mm]
&&
-\tau (4SX+S_x^2)F_d))
+2\eb
(m^2
(S_x(SV_2-XV_1-zQ^2S_p)+4M^2Q^2V_+) 
F_{2-}
\nonumber\\[1mm]
&&
+((Q^2+4m^2)(4M^2V_--zS_x^2)
+S_p(SV_2-XV_1))F_{IR})],
\nonumber\\[1mm]
\theta^0 _{53}&=&{\hat \theta^0 }_{53}+
\frac {\lambda_ eS}{\lambda_ 1\sqrt{\lambda_ S}}
[\ea (8m^2(\tau (\tau M^2-S_x )-Q^2)F_{21}+(Q^2(4\tau M^2+S_p)+2\tau SS_x)F_{1+}+\tau (4m^2(2\tau M^2-S_x)
\nonumber\\[1mm]&&
+Q^2(S_x-4S)-2\tau S^2)F_d)+2\eb(2m^2(S_x(2zQ^2+2V_-+(\tau z-\mu )S_x)-4M^2(\mu Q^2+\tau V_-))F_{21} 
\nonumber\\[1mm]
&&
+ \tau  (2m^2(zS_x^2-4M^2V_-)-2M^2Q^2V_1+S(z S_xQ^2-SV_2+X V_1))F_d)],
\nonumber\\[1mm]
{\hat \theta^0 }_{53}&=&\frac {2\lambda_ eS}{\lambda_ 1\sqrt{\lambda_ S}}F_{21}
[
\eb
(2(\mu Q^2+\tau V_1)(SX-M^2Q^2)
+(Q^2+\tau S)(zQ^2S_x-SV_2-XV_1))
-\ea (Q^2+\tau S )^2],
\nonumber\\[1mm]
\theta^0 _{62}&=&\frac 1{2\lambda_ 1}[
\ea
((4M^2Q^2(Q^2+4m^2)-S_x^2(Q^2-4m^2)-8Q^2SX)(S_xF_{1+}+2m^2F_{2-}-\tau S_pF_d)
\nonumber\\[1mm]
&&
+2S_p(2\tau (2M^2(Q^2+2m^2)-SX)-S_x(Q^2+4m^2) )F_{IR})
+2S_p\eb 
(m^2(S_x(zS_pQ^2-SV_2+V_1X)
\nonumber\\[1mm]
&&
-4M^2Q^2V_+)F_{2-} 
+((Q^2+4m^2)(zS_x^2-4M^2V_-)
+S_p(X V_1-S V_2))F_{IR})],
\nonumber\\[1mm]
\theta^0 _{63}&=&\frac 1{2\lambda_ 1}[2
\ea 
((2Q^2(SX-2M^2Q^2)-\tau S_x(S_x^2+3SX-4m^2M^2)-(Q^2+2m^2)S_x^2)F_{1+}
\nonumber\\[1mm]
&&
+m^2(2\tau(2M^2(Q^2+2m^2)-SX)-S_x(Q^2+4 m^2))F_{2-}
-Q^2S_pF_{IR}+S_p(\tau^2(S_x^2+2SX
\nonumber\\[1mm]
&&
-2M^2(Q^2+4m^2))+ 
2\tau S_x(Q^2+2m^2)-4m^2Q^2)F_d + S_p S_x^2)
+\eb
(
((Q^2+4m^2)(zS_x^2-4M^2V_-)
\nonumber\\[1mm]
&&
+S_p(XV_1-SV_2))
(S_xF_{1+}+2m^2F_{2-})
+2m^2(S_x(zQ^2S_p-SV_2+XV_1)-4M^2Q^2V_+)F_{2+} 
\nonumber\\[1mm]
&&
+(
4\tau (M^2Q^2(4SV_-+S_x(V_2-V_-))+2SX(SV_2-XV_1)
+2m^2S_p(4M^2V_--zSx^2))
\nonumber\\[1mm]
&&
+(3\tau S_x+2(Q^2-2m^2))(SV_2- 
XV_1-zQ^2S_p)S_x+8(Q^2-2m^2)M^2Q^2V_+)F_d)],
\nonumber\\[1mm]
\theta^0 _{64}&=&\frac 1{2\lambda_ 1}[
\ea
(((Q^2+4m^2)S_x+2\tau (SX-2M^2(Q^2+2m^2)))F_{1+}+S_p(\tau Q^2F_d-2S_x))
\nonumber\\[1mm]
&&
+\eb(((Q^2+4m^2)(4M^2V_--zS_x^2)+S_p(SV_2-XV_1))F_{1+}+\tau (4M^2Q^2V_++S_x(SV_2-XV_1-zQ^2S_p))F_d)],
\nonumber\\[1mm]
\theta^0 _{72}&=&
\frac {\lambda_eS}{2\sqrt{\lambda_S}}
[Q^2(4M^2V_--zS_x^2)F_{1+}
+m^2(\mu \lambda_Y-2S_x(zQ^2+V_-))F_{2-}+(2(4\tau M^2-S_x)V_+
\nonumber\\[1mm]
&&
+(\mu-2\tau z)S_pS_x- 
2SV_2+2XV_1)F_{IR}+ \tau (Q^2(zS_xS_p-4M^2V_+)+S_x(XV_1-SV_2))F_d],
\nonumber\\[1mm]
\theta^0 _{73}&=&
\frac {\lambda_eS}{4\sqrt{\lambda_S}}
[(S_x(4zQ^2+2V_--\mu S_x)-8\mu M^2Q^2)F_{1+}+2m^2 (4\mu \tau M^2+2V_--(\mu +2\tau z)S_x)F_{2-}
\nonumber\\[1mm]
&&
+2(2V_+-\mu S_p)F_{IR}+ 
\tau (4(S_x-2\tau M^2)V_++ S_p((2\tau z-\mu )S_x-2V_-))F_d],
\nonumber\\[1mm]    
\theta^0 _{74}&=&
\frac {\lambda_eS}{4\sqrt{\lambda_S}}
[((\mu +2\tau z)S_x-2V_{-}-4\mu \tau M^2)F_{1+}+\tau (\mu S_p-2V_{+})F_d],
\nonumber\\[1mm]
\theta^0 _{82}&=&\frac 1{\lambda_1}
[\ea(
(Q^2S_x(S_xV_+-2SV_2)-2V_-(2\lambda_1+Q^2SS_x))
F_{1+}
-2m^2(2\mu \lambda_1+Q^2S_pV_+)F_{2-}
\nonumber\\[1mm]
&&
+V_+(2m^2(2\tau(2(Q^2+2m^2)M^2
-SX)
-(Q^2+4m^2)S_x)F_{2+}
+(4m^2((3Q^2+4m^2)S_x
\nonumber\\[1mm]
&&
+\tau (2SX-4(3Q^2+2m^2)M^2-S_x^2))
+Q^2(\tau (12SX+S_x^2)
+2Q^2(S_x-6\tau M^2)))F_d))
\nonumber\\[1mm]
&&
+2V_+\eb (
((Q^2+4m^2)(zS_x^2-4M^2V_-)
+S_p(XV_1-SV_2))F_{IR}
+m^2(S_x(XV_1-SV_2+zS_pQ^2)
\nonumber\\[1mm]
&&
-4Q^2 V_+ M^2)F_{2-})],
\nonumber\\[1mm]
\theta^0 _{83}&=&\frac 1{2\lambda_1}
[
\ea
((2\mu (Q^2-2m^2)Q^2S_p+\tau(2(Q^2+8m^2)S_xV_+ +Q^2(\mu S_xS_p+2SV_1-2XV_2))
\nonumber\\[1mm]
&&
-2\tau ^2(4(Q^2+4m^2)V_+M^2-S_p(SV_1+XV_2)))F_d+ 
2\mu m^2(2\tau (2(Q^2+2m^2)M^2-SX)-(Q^2+4m^2)S_x)F_{2-}
\nonumber\\[1mm]
&&
+4S_pS_xV_--2\mu m^2Q^2 S_p F_{2+} + 
(2\tau (X^2V_2-S^2V_1)+8m^2V_-(2\tau M^2-S_x)+Q^2(4\mu(SX-2Q^2M^2)
\nonumber\\[1mm]
&&
-S_x(2V_-+\mu S_x)))F_{1+})
+2 \eb(((Q^2+4m^2)(zS_x^2-4M^2V_-)+S_p(XV_1-SV_2))(V_-F_{1+}+\mu m^2F_{2-})
\nonumber\\[1mm]
&&
+\mu m^2  (S_x(zS_pQ^2+XV_1-SV_2)- 
4Q^2V_+M^2)F_{2+}+(\mu (Q^2-2m^2)(4Q^2V_+M^2+S_x(SV_2-XV_1
\nonumber\\[1mm]
&&
-z S_pQ^2))+\tau(S_xS_pV_2(V_1+V_+)+ 
2V_-V_+((S_x-4S)X+2(3Q^2+8m^2)M^2)+zS_x(Q^2(XV_2-SV_1)
\nonumber\\[1mm]
&&
-(Q^2+8m^2)S_xV_+)))F_d)
],
\nonumber\\[1mm]
\theta^0 _{84}&=&\frac \mu {2\lambda_1}
[\ea (((Q^2+4m^2)S_x+2\tau (SX-2(Q^2+2m^2)M^2))F_{1+}+ S_p(\tau Q^2F_d-2S_x))
\nonumber\\[1mm]
&&
+\eb(((Q^2+4m^2)(4M^2V_--zS_x^2)+S_p(SV_2-XV_1))F_{1+}+\tau (4Q^2V_+M^2+S_x(SV_2-XV_1
\nonumber\\[1mm]
&&
-zS_pQ^2))F_d)],
\nonumber\\[1mm]
\theta^0 _{91}&=&\frac{2\lambda_e S }{\sqrt{\lambda_S}} (Q^2(zS_xV_+-m_h^2S_p)+V_- (SV_2-XV_1))F_{IR},
\nonumber\\[1mm]
\theta^0 _{92}&=&\frac {\lambda_e S}{\sqrt{\lambda_S}}
(Q^2S_x(zV_--m_h^2)F_{1+}+ 
 m^2 (Q^2(\mu z S_x-2m_h^2)+V_- (\mu S_x-2V_-))F_{2-} 
\nonumber\\[1mm]
&&
+\tau (Q^2(m_h^2S_p-zS_xV_+)+V_-(XV_1-SV_2))F_d+ 
(2V_-(2\mu S-V_+ )+2\tau (zS_xV_+-m_h^2S_p)
\nonumber\\[1mm]
&&
-\mu (V_1+V_-)S_x)F_{IR}),
\nonumber\\[1mm]
\theta^0 _{93}&=&\frac {\lambda_e S}{2\sqrt{\lambda_S}}
((2(2m_h^2Q^2+V_-^2)-\mu S_x(2zQ^2+V_-))F_{1+}+m^2(\mu ((2\tau z-\mu )S_x+2V_-)-4\tau m_h^2)F_{2-} 
\nonumber\\[1mm]
&&
+\mu (2V_+-\mu S_p)F_{IR}+\tau (2\tau (m_h^2S_p-zS_xV_+)+\mu S_x(V_-+V_1)+2V_-(V_+-2\mu S))F_d),
\nonumber\\[1mm]
\theta^0 _{94}&=&\frac {\lambda_eS}{4\sqrt{\lambda_S}}
((2\tau (2m_h^2-\mu zS_x)+\mu (\mu S_x-2V_-))F_{1+}+\mu \tau (\mu S_p-2V_+) F_d).
\label{theta0}
\end{eqnarray}
The quantities $\theta^1_{ij}$ have the form
\begin{eqnarray}
\theta^1_{51}&=&0,
\nonumber\\[1mm]
\theta^1_{52}&=&\frac{2 m^2\lambda_e }{\lambda_1\sqrt{\lambda_S}}
[\ea (2(2m^2\lambda_Y+(Q^2+\tau S)(2M^2Q^2+ SS_x))F_{21}
-S_x\lambda_YF_{1+}
+(2Q^2XS_x
+\tau S_x(2S^2-S_p^2)
\nonumber\\[1mm]
&&
+4M^2Q^2(\tau S-Q^2)-4m^2\lambda_Y)F_d)
+2\eb 
(S_x(zS_pQ^2+XV_1-SV_2)-4M^2Q^2V_p)
(XF_d-SF_{21})
],
\nonumber\\[1mm]
\theta^1_{53}&=&\frac{2 m^2\lambda_e}{\lambda_1\sqrt{\lambda_S}}
[\ea
(
2((Q^2+2m^2)(2\tau M^2+X)-(\tau X+2m^2)S)F_{21}
-\lambda_YF_{1+}
+(4m^2(S_x-2\tau M^2)
+2Q^2S
\nonumber\\[1mm]
&&
+\tau (S^2+X^2))F_d)
+2\eb
((2m^2(zS_x^2-4M^2V_-)+2M^2Q^2V_2
+X(XV_1-SV_2-zS_xQ^2))F_{21}
\nonumber\\[1mm]
&&
+(2m^2(4M^2V_--zS_x^2)+2M^2Q^2V_1+ 
S(SV_2-XV_1-zQ^2S_x))F_d)
],
\nonumber\\[1mm]
\theta^1_{71}&=&0,
\nonumber\\[1mm]
\theta^1_{72}&=&\frac{m^2\lambda_e }{\sqrt{\lambda_S}}
((4M^2(\tau SV_--Q^2V_+)-S_x^2(\tau zS+z Q^2+V_1)+\mu  \lambda_Y  S)F_{21}
+(4M^2(Q^2V_+-\tau XV_-)
\nonumber\\[1mm]
&&
+S_x^2(\tau z X+V_2-z Q^2)-\mu \lambda_Y  X)F_d),
\nonumber\\[1mm]
\theta^1_{73}&=&\frac{m^2\lambda_e }{\sqrt{\lambda_S}}
[(2M^2(\mu (Q^2+\tau S)-2\tau V_+)+S_x((\tau z-2\mu )S+2V_+-zQ^2)+(\mu-\tau z )S_x^2)F_{21}
\nonumber\\[1mm]
&&
+(
2M^2(\mu (Q^2-\tau X)+2\tau V_+)
+       
S_x((2\mu-\tau z )S-zQ^2-2V_+)
-\mu S_x^2)F_d],
\nonumber\\[1mm]
\theta^1_{74}&=&\frac{m^2\lambda_e }{\sqrt{\lambda_S}}
[(2\mu \tau M^2+\mu X-\tau zS_x-V_2)F_{21}+(2\mu \tau M^2+V_1-\mu S-\tau zS_x)F_d],
\nonumber\\[1mm]
\theta^1_{91}&=&\frac{4\lambda_e m^2(m_h^2\lambda_Y+4M^2V_-^2-zS_x^2(zQ^2+2V_-))}{\sqrt{\lambda_S}}F_{IR},
\nonumber\\[1mm]
\theta^1_{92}&=&\frac{2m^2\lambda_e}{\sqrt{\lambda_S}}
[2m^2(2m_h^2(2\tau M^2-S_x)+2(zS_x-2\mu M^2)V_-+z(\mu-\tau z)S_x^2) F_{2+}
\nonumber\\[1mm]
&&
+S_x(zQ^2(\mu S-V_+)+V_-((\mu +\tau z)S-V_1)-m_h^2(Q^2+\tau S))F_{21}
\nonumber\\[1mm]
&&
+(S_x((m_h^2-zV_-)(\tau X+3Q^2+8m^2)+(V_-+zQ^2)(V_2-\mu X))
\nonumber\\[1mm]
&&
+2(4M^2(\mu V_--\tau m_h^2)+(z\tau-\mu )zS_x ^2)(Q^2+2m^2))F_d],
\nonumber\\[1mm]
\theta^1_{93}&=&\frac{m^2\lambda_e }{\sqrt{\lambda_S}}
[4m^2 (m_h^2+\mu ^2M^2-\mu zS_x)F_{2+}+(2m_h^2(\tau X-Q^2)+S_x(\mu (\tau z-\mu )S-2\tau zV_+ +\mu z Q^2
\nonumber\\[1mm]
&&
+\mu V_1)+2V_-(V_2-\mu X))F_{21}+(\mu ((Q^2+2m^2)(5zS_x-4\mu M^2)+(\mu-\tau z)XS_x)
\nonumber\\[1mm]
&&
-2m_h^2(\tau S+3Q^2+4m^2)+S_x(2\tau zV_+-\mu V_2-2\mu z m^2)+2V_-(\mu S-V_1))F_d],
\nonumber\\[1mm]
\theta^1_{94}&=&\frac{m^2\lambda_e }{\sqrt{\lambda_S}}
[(\mu (\tau zS_x+\mu X-V_2)-2\tau m_h^2)F_{21}+(\mu (\tau zS_x+V_1-\mu S)-2\tau m_h^2))F_d].
\end{eqnarray}

\end{widetext}
The variable $\mu $ is defined as
\begin{eqnarray}
\mu&=&\frac{kp_h}{kp}=\frac{p_{h0}}{M}+\frac{p_l(2\tau M^2-S_x)}{M\sqrt{\lambda_Y}}
\nonumber\\&&
-2Mp_t\cos(\phi_h+\phi_k)\sqrt{\frac{(\tau_{\rm max}-\tau)(\tau-\tau_{\rm min})}{\lambda_Y}}
\nonumber\\
\label{mudef}
\end{eqnarray}

The quantities $F_i$ ($i=d,1+,2+,2-,IR$) are expressed through
\begin{eqnarray}
z_1&=&\frac {k_1k}{pk}
\nonumber\\
&=&\frac {Q^2S_p+\tau (SS_x+2M^2Q^2)-2M\sqrt{\lambda_z}\cos \phi_k}{\lambda_Y},
\nonumber\\[1mm]
z_2&=&\frac {k_1k}{pk}
\nonumber\\
&=&\frac {Q^2S_p+\tau (XS_x-2M^2Q^2)-2M\sqrt{\lambda_z}\cos \phi_k}{\lambda_Y},
\nonumber\\[1mm]
\lambda_z&=&(\tau _{\rm max}-\tau)(\tau-\tau _{\rm min})\lambda_ 1
\nonumber\\[1mm]
\label{z1z2}
\end{eqnarray}
in the following way:
\begin{eqnarray}
F_{2\pm}&=&F_{22}\pm F_{21}=\frac 1{z_2^2}\pm \frac 1{z_1^2},
\nonumber\\[1mm]
F_d&=&\frac 1{z_1 z_2},
\nonumber\\[1mm]
F_{1+}&=&\frac 1{z_1}+\frac 1{z_2},
\nonumber\\[1mm]
F_{IR}&=&m^2F_{2+}-(Q^2+2m^2) F_d.
\label{fff}
\end{eqnarray}

\section{Calculation of $\delta_S$ and $\delta_H$}
\label{ap2}
The dimensional regularization is used for the calculation of $\delta_S$ in (\ref{dsh}), 
\begin{eqnarray}
\frac{d^3k^\prime }{k_0^\prime}&\to& \frac{d^{n-1}k^\prime}{(2\pi \nu)^{n-4}k_0^\prime}
\nonumber\\
&=&
\frac{2\pi ^{n/2-1}k_0^{\prime n-3}dk_0^\prime(1-x^2)^{n/2-2} dx }{(2\pi \nu)^{n-4}\Gamma(n/2-1)},
\label{dnk}
\end{eqnarray}
where $x=\cos \theta $ [$\theta$ is defined as the spatial angle between the photon three-momentum and ${\bf k}_i^\prime $ ($i=1-3$ ) 
that are introduced below]
and $\nu $ is an arbitrary parameter of the dimension of a mass. The Feynman parametrization of propagators in 
${F_{IR}}$,
\begin{eqnarray}
F_{IR}&=&
\frac {R^2}  {4k_0^{\prime 2}}
\int\limits_0^1 dy
{\mathcal F}(x,y),
\label{firxy}
\end{eqnarray}
where $y$ is the Feynman parameter and
\begin{eqnarray}
{\mathcal F}(x,y)
&=&
\frac {m^2}{k_{10}^{\prime 2}(1-x\beta_1)^2}+
\frac {m^2}{k_{20}^{\prime 2}(1-x\beta_2)^2}
\nonumber\\&&
-\frac {Q_m^2}{k_{30}^{\prime 2}(1-x\beta_3)^2}.
\label{fxy}
\end{eqnarray}
The energies of the real photon ($k_{0}^{\prime }$), initial ($k_{10}^{\prime }$) 
and final ($k_{20}^{\prime }$) leptons are defined in the system 
${\bf p}+{\bf q}-{\bf p}_h=0$ while $k_{30}^{\prime }=yk_{10}^{\prime }+(1-y)k_{20}^{\prime }$
and $\beta _i=|{\bf k}^{\prime }_i|/k_{i0}^{\prime }$.

Then, the substitution of Eqs.~(\ref{dnk}) and (\ref{fxy}) into the definitions of $\delta_S$ by Eq.~(\ref{dsh}),
the integration over $k_0^{\prime }$, and expansion of the obtained result into the Laurent series around
$n=4$ result in
\begin{eqnarray}
\delta_S&=&\delta_S^{IR}+\delta_S^1,
\end{eqnarray}
where
\begin{eqnarray}
\delta_S^{IR}&=&-\frac 12\biggl[P_{IR}+\log \frac{{\bar k}_0}{\nu}\biggr] 
\int\limits_0^1dy
\int\limits_{-1}^1dx
{\mathcal F}(x,y)
\end{eqnarray}
 and
\begin{eqnarray}
\delta_S^1&=&-\frac 14
\int\limits_0^1dy
\int\limits_{-1}^1dx\log (1-x^2)
{\mathcal F}(x,y).
\end{eqnarray}
Here $P_{IR}$ is the infrared divergent term defined by Eq.~(\ref{pir}).
Since $k_{30}^{\prime 2}-|{\bf k}^\prime_3|^2=m^2+y(1-y)Q^2$,
the integration over $x$ and $y$ variables in $\delta_S^{IR}$ is performed explicitly:
\begin{eqnarray}
\delta_S^{IR}&=&2(Q^2_mL_m-1)\biggl[P_{IR}+\log \frac{{\bar k}_0}{\nu}\biggr].
\end{eqnarray}

For the covariant
analytical integration in $\delta_S^1$ we express
the initial and final lepton energies through the invariants:
\begin{eqnarray}
k_{10}^\prime=\frac{S^\prime }{2\sqrt{p_x^2}},\qquad
k_{20}^\prime=\frac{X^\prime }{2\sqrt{p_x^2}}.
\end{eqnarray}
As a result,
\begin{eqnarray}
\delta_S^1
&=&2(Q^2_mL_m-1)\log (2)+\frac 12S^\prime L_{S^\prime}+\frac 12X^\prime L_{X^\prime}
\nonumber\\&&
+S_\phi,
\end{eqnarray}
where the quantities $L_m$, $L_{S^\prime}$, and $L_{X^\prime}$ are
\begin{eqnarray}
L_m&=&\frac 1{\sqrt{\lambda_m}}\log\frac {\sqrt{\lambda_m}+Q^2}{\sqrt{\lambda_m}-Q^2},
\nonumber \\
L_{S^\prime}&=&\frac 1{\sqrt{\lambda_S^\prime}}\log\frac{S^\prime+\sqrt{\lambda_S^\prime}}{S^\prime-\sqrt{\lambda_S^\prime}},
\nonumber\\[1mm]
L_{X^\prime}&=&\frac 1{\sqrt{\lambda_X^\prime}}\log\frac{X^\prime+\sqrt{\lambda_X^\prime}}{X^\prime-\sqrt{\lambda_X^\prime}}
\label{lms}
\end{eqnarray}
and
\begin{eqnarray}
S_\phi&=&\frac 1 2Q_m^2\int\limits_0^1\frac{dy}{\beta_3(m^2+y(1-y)Q^2)}
\log \frac{1-\beta_3}{1+\beta_3}.
\nonumber\\
\end{eqnarray}
The explicit expression for $S_\phi$ after integration over $y$ is given in Eq.~(\ref{sph0}).

For the calculation of $\delta_H$ we carry out the integration
in the same reference system ${\bf p}+{\bf q}-{\bf p}_h=0$,
\begin{eqnarray}
\delta_H&=&-\frac 1\pi\int\limits_{{\bar k}_0}^{k_0^{\rm max \prime}} k_0^\prime dk_0^\prime
\int\limits_0^{\pi}\sin(\theta_k^\prime )d\theta_k^\prime 
\int\limits_0^{2\pi}d\phi_k^\prime 
\frac{F_{IR}}{R^2},
\nonumber\\
\label{dh}
\end{eqnarray}
where $\theta_k^\prime$ is the angle between $\bf k$ and $\bf q$ three momenta, and  
$\phi_k^\prime$ is the angle between $({\bf k}_1,{\bf k}_2)$ and $({\bf k},{\bf q})$ planes.

In this system 
\begin{eqnarray}
z_1&=&\frac{2k_0^\prime}R(k_{10}^\prime-k_t^\prime \cos\phi_k^\prime\sin\theta_k^\prime-k_{13}^\prime\cos\theta_k^\prime),
\nonumber\\
z_2&=&\frac{2k_0^\prime}R(k_{20}^\prime-k_t^\prime \cos\phi_k^\prime\sin\theta_k^\prime-k_{23}^\prime\cos\theta_k^\prime),
\end{eqnarray}
which allows us to take the first integration in respect to $\phi _k$:
\begin{eqnarray}
\delta_H&=&\int\limits_{{\bar k}_0}^{k_0^{ max \prime}} \frac{ dk_0^\prime}{2 k_0^\prime}
\int\limits_0^{\pi}\sin(\theta_k^\prime )d\theta_k^\prime 
\Biggl[\frac{Q^2_m}{B_1-B_2}
\Biggl(\frac 1{\sqrt{C_2}}
\nonumber\\&& 
-\frac 1{\sqrt{C_1}}\Biggr)
-\frac{m^2 B_1}{C^{3/2}_1}
-\frac{m^2 B_2}{C^{3/2}_2}
\Biggr].
\end{eqnarray}
Here
\begin{eqnarray}
B_i=k_{i0}^\prime-\cos(\theta _k^\prime) k_{i3}^\prime,\;
C_i=B^2_i-\sin^2(\theta _k^\prime) k_{t}^{\prime 2}\;
\end{eqnarray}
for $i=1,2$.

After the integration with respect to $\theta _k^\prime$
and the use of the following replacements:
\begin{eqnarray}
k_t^\prime &=&\sqrt{k_{10}^{\prime 2}-k_{13}^{\prime 2}-m^2}=\sqrt{k_{20}^{\prime 2}-k_{23}^{\prime 2}-m^2},
\nonumber\\[2mm]
k_{13}^\prime&=&\frac{2k_{10}^\prime q_0^\prime+Q^2}{2\sqrt{Q^2+q_0^{\prime 2}}},\;
k_{23}^\prime=\frac{2k_{20}^\prime q_0^\prime-Q^2}{2\sqrt{Q^2+q_0^{\prime 2}}}
\end{eqnarray}
\vspace*{1mm}

\noindent
with $q_0^\prime=k^\prime_{10}-k^\prime_{20}$, 
the hard contribution $\delta_H$
is expressed in the form
\begin{eqnarray}
\delta_H&=&2\int\limits_{{\bar k}_0}^{k_0^{\rm max \prime}} \frac{ dk_0}{k_0}
(Q^2_mL_m-1).
\end{eqnarray}

Since
$k_0^{\rm max \prime}=(p_x^2-M_{th}^2)/2\sqrt{p_x^2}$
the integration for $\delta_H$ is finally presented in the form of (\ref{ird}).

\section{Calculation of $\Lambda_\mu$ and $\Pi^l_{\alpha \mu}$}
\label{ap3}
The $\gamma$-matrix recombination, convolution over $\alpha$ indexes
 in Eq.~(\ref{lp0}) for $\Lambda_{\mu }$ 
and calculation of the traces for $\Pi^l_{\alpha \mu }$ in $n$-dimensional space result in
\begin{eqnarray}
\Lambda_{\mu }&=&\frac{\alpha}{4\pi}\biggl\{
\gamma_\mu [(n-2)J^\delta_{\delta}
-4J^\delta(k_{1\delta}+k_{2\delta})
\nonumber\\&&
+2Q_m^2J
]
+2\gamma_\delta [ 
2J^\delta(k_{1\mu}+k_{2\mu})
\nonumber\\&&
-(n-2) J^\delta_{\mu}
]
-4mJ_{\mu }
\biggr\},
\nonumber \\
\Pi^l_{\alpha \mu }&=&\frac{\alpha}{\pi}\frac 1{Q^2}\biggl(
\sum_{i=e,\mu,\tau}
\biggl\{
g_{\alpha \mu }(
q_\delta J^\delta_{i}
+m^2_iJ_{i}
-J^\delta_{i\delta}
)
\nonumber\\&&
+2J_{i\alpha \mu}
-q_\alpha J_{i\mu }
-q_\mu J_{i\alpha }
\biggr\}
\biggr)
, 
\label{lp1}
\end{eqnarray}
where
\begin{widetext}
\begin{eqnarray}
J=\frac 1{i\pi ^2}\lim\limits_{n\to 4}\int \frac{(2\pi \nu)^{4-n}d^nl}{l^2(l^2-2lk_2)(l^2-2lk_1)}
&=&-2L_m\biggl (P_{IR}+\log\frac m\nu \biggr )-\frac 12\sqrt{\lambda_m}L_m^2+\frac 1{2\sqrt{\lambda_m}}
\biggl(\pi^2
-4{\rm Li}_2 \frac{2\sqrt{\lambda_m}}{\sqrt{\lambda_m}+Q^2}\biggr),
\nonumber\\
J_\delta=\frac 1{i\pi ^2}\lim\limits_{n\to 4}\int \frac{l_{\delta}(2\pi \nu)^{4-n}d^nl}{l^2(l^2-2lk_2)(l^2-2lk_1)}
&=&-L_m(k_{1\delta}+k_{2\delta}),
\nonumber\\
J_{\delta\rho }=\frac 1{i\pi ^2}\lim\limits_{n\to 4}\int \frac{l_{\delta}l_{\rho}(2\pi \nu)^{4-n}d^nl}{l^2(l^2-2lk_2)(l^2-2lk_1)}
&=&\frac 14\biggl\{ g_{\delta\rho }
\biggl(3-2P_{UV}-2\log\frac m \nu -\frac{\lambda_m}{Q^2}L_m
\biggr)
+q_\delta q_\rho\frac {2Q^2-\lambda_mL_m}{Q^4}
\nonumber\\&&
-L_m(k_{1\delta}+k_{2\delta})(k_{1\rho}+k_{2\rho})
\biggr\}, 
\nonumber\\
J_i=\frac 1{i\pi ^2}\lim\limits_{n\to 4}\int \frac{(2\pi \nu)^{4-n}d^nl}{(l^2-m_i^2)((l-q)^2-m_i^2)}
&=&2-2P_{UV}-2\log\frac {m_i} \nu -\frac {\lambda^i_m}{Q^2}L^i_m,
\nonumber\\
J_{i\delta}=\frac 1{i\pi ^2}\lim\limits_{n\to 4}\int \frac{l_{\delta}(2\pi \nu)^{4-n}d^nl}{(l^2-m_i^2)((l-q)^2-m_i^2)}
&=&\frac 1 2q_\delta
J_i,
\nonumber\\
J_{i\delta\rho }=\frac 1{i\pi ^2}\lim\limits_{n\to 4}\int \frac{l_{\delta}l_{\rho}(2\pi \nu)^{4-n}d^nl}{(l^2-m_i^2)((l-q)^2-m_i^2)}
&=&\frac 1{72}\biggl\{ g_{\delta\rho }
\biggl(6\biggl[Q^2-\frac {3\lambda_m^i}{Q^2}\biggr](P_{UV}+\log\frac {m_i} \nu )
+\biggl[21
-\frac{6\lambda_m^i}{Q^2}L_m^i\biggr]\frac{\lambda_m^i}{Q^2}
-5Q^2\biggr)
\nonumber\\&&
+q_\delta q_\rho
\biggl(
40-48P_{UV}-48\log\frac {m_i}\nu +12\frac{\lambda_m^i}{Q^4}-6\frac{\lambda_m^i}{Q^2}\biggl[3+\frac{\lambda_m^i}{Q^4}\biggr]L_m
\biggr)
\biggr\}. 
\label{intv}
\end{eqnarray}
\end{widetext}
The infrared divergent $P_{IR}$ term is defined by Eq.~(\ref{pir}) while the ultraviolet divergent term has the same structure 
$P_{UV}=P_{IR}$ and
\begin{eqnarray}
&&
L_m^i=\frac 1{\sqrt{\lambda_m^i}}\log\frac {\sqrt{\lambda_m^i}+Q^2}{\sqrt{\lambda_m^i}-Q^2},\;
\lambda_m^i=Q^2(Q^2+4m_i^2).
\nonumber \\
\label{lmi}
\end{eqnarray}
After substituting (\ref{intv}) into (\ref{lp1}) and using $nP_{UV}=4P_{UV}+1+{\mathcal O}(n-4)$ we find the final expressions for $\Lambda_\mu $ and 
$\Pi^l_{\alpha \mu }$  
 (\ref{lp2}).
\bibliography{haprad}

\begin{thebibliography}{21}%
\makeatletter
\providecommand \@ifxundefined [1]{%
 \@ifx{#1\undefined}
}%
\providecommand \@ifnum [1]{%
 \ifnum #1\expandafter \@firstoftwo
 \else \expandafter \@secondoftwo
 \fi
}%
\providecommand \@ifx [1]{%
 \ifx #1\expandafter \@firstoftwo
 \else \expandafter \@secondoftwo
 \fi
}%
\providecommand \natexlab [1]{#1}%
\providecommand \enquote  [1]{``#1''}%
\providecommand \bibnamefont  [1]{#1}%
\providecommand \bibfnamefont [1]{#1}%
\providecommand \citenamefont [1]{#1}%
\providecommand \href@noop [0]{\@secondoftwo}%
\providecommand \href [0]{\begingroup \@sanitize@url \@href}%
\providecommand \@href[1]{\@@startlink{#1}\@@href}%
\providecommand \@@href[1]{\endgroup#1\@@endlink}%
\providecommand \@sanitize@url [0]{\catcode `\\12\catcode `\$12\catcode
  `\&12\catcode `\#12\catcode `\^12\catcode `\_12\catcode `\%12\relax}%
\providecommand \@@startlink[1]{}%
\providecommand \@@endlink[0]{}%
\providecommand \url  [0]{\begingroup\@sanitize@url \@url }%
\providecommand \@url [1]{\endgroup\@href {#1}{\urlprefix }}%
\providecommand \urlprefix  [0]{URL }%
\providecommand \Eprint [0]{\href }%
\providecommand \doibase [0]{http://dx.doi.org/}%
\providecommand \selectlanguage [0]{\@gobble}%
\providecommand \bibinfo  [0]{\@secondoftwo}%
\providecommand \bibfield  [0]{\@secondoftwo}%
\providecommand \translation [1]{[#1]}%
\providecommand \BibitemOpen [0]{}%
\providecommand \bibitemStop [0]{}%
\providecommand \bibitemNoStop [0]{.\EOS\space}%
\providecommand \EOS [0]{\spacefactor3000\relax}%
\providecommand \BibitemShut  [1]{\csname bibitem#1\endcsname}%
\let\auto@bib@innerbib\@empty
\bibitem [{\citenamefont {Barone}\ \emph {et~al.}(2002)\citenamefont {Barone},
  \citenamefont {Drago},\ and\ \citenamefont {Ratcliffe}}]{Barone}%
  \BibitemOpen
  \bibfield  {author} {\bibinfo {author} {\bibfnamefont {V.}~\bibnamefont
  {Barone}}, \bibinfo {author} {\bibfnamefont {A.}~\bibnamefont {Drago}}, \
  and\ \bibinfo {author} {\bibfnamefont {P.~G.}\ \bibnamefont {Ratcliffe}},\
  }\href {\doibase 10.1016/S0370-1573(01)00051-5} {\bibfield  {journal}
  {\bibinfo  {journal} {Phys. Rept.}\ }\textbf {\bibinfo {volume} {359}},\
  \bibinfo {pages} {1} (\bibinfo {year} {2002})}\BibitemShut {NoStop}%
\bibitem [{\citenamefont {Airapetian}\ \emph {et~al.}(2005)\citenamefont
  {Airapetian} \emph {et~al.}}]{HERMES}%
  \BibitemOpen
  \bibfield  {author} {\bibinfo {author} {\bibfnamefont {A.}~\bibnamefont
  {Airapetian}} \emph {et~al.} (\bibinfo {collaboration} {HERMES}),\ }\href
  {\doibase 10.1103/PhysRevLett.94.012002} {\bibfield  {journal} {\bibinfo
  {journal} {Phys. Rev. Lett.}\ }\textbf {\bibinfo {volume} {94}},\ \bibinfo
  {pages} {012002} (\bibinfo {year} {2005})}\BibitemShut {NoStop}%
\bibitem [{\citenamefont {Alexakhin}\ \emph {et~al.}(2005)\citenamefont
  {Alexakhin} \emph {et~al.}}]{COMPASS}%
  \BibitemOpen
  \bibfield  {author} {\bibinfo {author} {\bibfnamefont {V.~{\relax Yu}.}\
  \bibnamefont {Alexakhin}} \emph {et~al.} (\bibinfo {collaboration}
  {COMPASS}),\ }\href {\doibase 10.1103/PhysRevLett.94.202002} {\bibfield
  {journal} {\bibinfo  {journal} {Phys. Rev. Lett.}\ }\textbf {\bibinfo
  {volume} {94}},\ \bibinfo {pages} {202002} (\bibinfo {year}
  {2005})}\BibitemShut {NoStop}%
\bibitem [{\citenamefont {Qian}\ \emph {et~al.}(2011)\citenamefont {Qian} \emph
  {et~al.}}]{JLab1}%
  \BibitemOpen
  \bibfield  {author} {\bibinfo {author} {\bibfnamefont {X.}~\bibnamefont
  {Qian}} \emph {et~al.} (\bibinfo {collaboration} {Jefferson Lab Hall A}),\
  }\href {\doibase 10.1103/PhysRevLett.107.072003} {\bibfield  {journal}
  {\bibinfo  {journal} {Phys. Rev. Lett.}\ }\textbf {\bibinfo {volume} {107}},\
  \bibinfo {pages} {072003} (\bibinfo {year} {2011})}\BibitemShut {NoStop}%
\bibitem [{\citenamefont {Dudek}\ \emph {et~al.}(2012)\citenamefont {Dudek}
  \emph {et~al.}}]{12Jlab}%
  \BibitemOpen
  \bibfield  {author} {\bibinfo {author} {\bibfnamefont {J.}~\bibnamefont
  {Dudek}} \emph {et~al.},\ }\href {\doibase 10.1140/epja/i2012-12187-1}
  {\bibfield  {journal} {\bibinfo  {journal} {Eur. Phys. J.}\ }\textbf
  {\bibinfo {volume} {A48}},\ \bibinfo {pages} {187} (\bibinfo {year}
  {2012})}\BibitemShut {NoStop}%
\bibitem [{\citenamefont {Akushevich}\ \emph {et~al.}(1997)\citenamefont
  {Akushevich}, \citenamefont {Ilyichev}, \citenamefont {Shumeiko},
  \citenamefont {Soroko},\ and\ \citenamefont {Tolkachev}}]{Polrad}%
  \BibitemOpen
  \bibfield  {author} {\bibinfo {author} {\bibfnamefont {I.}~\bibnamefont
  {Akushevich}}, \bibinfo {author} {\bibfnamefont {A.}~\bibnamefont
  {Ilyichev}}, \bibinfo {author} {\bibfnamefont {N.}~\bibnamefont {Shumeiko}},
  \bibinfo {author} {\bibfnamefont {A.}~\bibnamefont {Soroko}}, \ and\ \bibinfo
  {author} {\bibfnamefont {A.}~\bibnamefont {Tolkachev}},\ }\href {\doibase
  10.1016/S0010-4655(97)00062-3} {\bibfield  {journal} {\bibinfo  {journal}
  {Comput. Phys. Commun.}\ }\textbf {\bibinfo {volume} {104}},\ \bibinfo
  {pages} {201} (\bibinfo {year} {1997})}\BibitemShut {NoStop}%
\bibitem [{\citenamefont {Soroko}\ and\ \citenamefont {Shumeiko}(1989)}]{SSh1}%
  \BibitemOpen
  \bibfield  {author} {\bibinfo {author} {\bibfnamefont {A.~V.}\ \bibnamefont
  {Soroko}}\ and\ \bibinfo {author} {\bibfnamefont {N.~M.}\ \bibnamefont
  {Shumeiko}},\ }\href@noop {} {\bibfield  {journal} {\bibinfo  {journal} {Sov.
  J. Nucl. Phys.}\ }\textbf {\bibinfo {volume} {49}},\ \bibinfo {pages} {838}
  (\bibinfo {year} {1989})},\ \bibinfo {note} {[Yad.
  Fiz.49,1348(1989)]}\BibitemShut {NoStop}%
\bibitem [{\citenamefont {Soroko}\ and\ \citenamefont {Shumeiko}(1991)}]{SSh2}%
  \BibitemOpen
  \bibfield  {author} {\bibinfo {author} {\bibfnamefont {A.~V.}\ \bibnamefont
  {Soroko}}\ and\ \bibinfo {author} {\bibfnamefont {N.~M.}\ \bibnamefont
  {Shumeiko}},\ }\href@noop {} {\bibfield  {journal} {\bibinfo  {journal} {Sov.
  J. Nucl. Phys.}\ }\textbf {\bibinfo {volume} {53}},\ \bibinfo {pages} {628}
  (\bibinfo {year} {1991})},\ \bibinfo {note} {[Yad.
  Fiz.53,1015(1991)]}\BibitemShut {NoStop}%
\bibitem [{\citenamefont {Akushevich}\ \emph {et~al.}(1999)\citenamefont
  {Akushevich}, \citenamefont {Shumeiko},\ and\ \citenamefont {Soroko}}]{ASSh}%
  \BibitemOpen
  \bibfield  {author} {\bibinfo {author} {\bibfnamefont {I.}~\bibnamefont
  {Akushevich}}, \bibinfo {author} {\bibfnamefont {N.}~\bibnamefont
  {Shumeiko}}, \ and\ \bibinfo {author} {\bibfnamefont {A.}~\bibnamefont
  {Soroko}},\ }\href {\doibase 10.1007/s100520050606, 10.1007/s100529900172}
  {\bibfield  {journal} {\bibinfo  {journal} {Eur. Phys. J.}\ }\textbf
  {\bibinfo {volume} {C10}},\ \bibinfo {pages} {681} (\bibinfo {year}
  {1999})}\BibitemShut {NoStop}%
\bibitem [{\citenamefont {Akushevich}\ \emph {et~al.}(2009)\citenamefont
  {Akushevich}, \citenamefont {Ilyichev},\ and\ \citenamefont
  {Osipenko}}]{AIO}%
  \BibitemOpen
  \bibfield  {author} {\bibinfo {author} {\bibfnamefont {I.}~\bibnamefont
  {Akushevich}}, \bibinfo {author} {\bibfnamefont {A.}~\bibnamefont
  {Ilyichev}}, \ and\ \bibinfo {author} {\bibfnamefont {M.}~\bibnamefont
  {Osipenko}},\ }\href {\doibase 10.1016/j.physletb.2008.12.058} {\bibfield
  {journal} {\bibinfo  {journal} {Phys. Lett.}\ }\textbf {\bibinfo {volume}
  {B672}},\ \bibinfo {pages} {35} (\bibinfo {year} {2009})}\BibitemShut
  {NoStop}%
\bibitem [{\citenamefont {Drechsel}\ \emph {et~al.}(1999)\citenamefont
  {Drechsel}, \citenamefont {Hanstein}, \citenamefont {Kamalov},\ and\
  \citenamefont {Tiator}}]{Maid}%
  \BibitemOpen
  \bibfield  {author} {\bibinfo {author} {\bibfnamefont {D.}~\bibnamefont
  {Drechsel}}, \bibinfo {author} {\bibfnamefont {O.}~\bibnamefont {Hanstein}},
  \bibinfo {author} {\bibfnamefont {S.~S.}\ \bibnamefont {Kamalov}}, \ and\
  \bibinfo {author} {\bibfnamefont {L.}~\bibnamefont {Tiator}},\ }\href
  {\doibase 10.1016/S0375-9474(98)00572-7} {\bibfield  {journal} {\bibinfo
  {journal} {Nucl. Phys.}\ }\textbf {\bibinfo {volume} {A645}},\ \bibinfo
  {pages} {145} (\bibinfo {year} {1999})}\BibitemShut {NoStop}%
\bibitem [{\citenamefont {Bardin}\ and\ \citenamefont {Shumeiko}(1977)}]{BSh}%
  \BibitemOpen
  \bibfield  {author} {\bibinfo {author} {\bibfnamefont {D.~{\relax Yu}.}\
  \bibnamefont {Bardin}}\ and\ \bibinfo {author} {\bibfnamefont {N.~M.}\
  \bibnamefont {Shumeiko}},\ }\href {\doibase 10.1016/0550-3213(77)90213-9}
  {\bibfield  {journal} {\bibinfo  {journal} {Nucl. Phys.}\ }\textbf {\bibinfo
  {volume} {B127}},\ \bibinfo {pages} {242} (\bibinfo {year}
  {1977})}\BibitemShut {NoStop}%
\bibitem [{\citenamefont {Mo}\ and\ \citenamefont {Tsai}(1969)}]{Mo-Tsai}%
  \BibitemOpen
  \bibfield  {author} {\bibinfo {author} {\bibfnamefont {L.~W.}\ \bibnamefont
  {Mo}}\ and\ \bibinfo {author} {\bibfnamefont {Y.-S.}\ \bibnamefont {Tsai}},\
  }\href {\doibase 10.1103/RevModPhys.41.205} {\bibfield  {journal} {\bibinfo
  {journal} {Rev. Mod. Phys.}\ }\textbf {\bibinfo {volume} {41}},\ \bibinfo
  {pages} {205} (\bibinfo {year} {1969})}\BibitemShut {NoStop}%
\bibitem [{\citenamefont {Tsai}(1971)}]{Tsai}%
  \BibitemOpen
  \bibfield  {author} {\bibinfo {author} {\bibfnamefont {Y.-S.}\ \bibnamefont
  {Tsai}}\ }(\bibinfo {year} {1971})\BibitemShut {NoStop}%
, SLAC-PUB-0848
\bibitem [{\citenamefont {Bacchetta}\ \emph {et~al.}(2004)\citenamefont
  {Bacchetta}, \citenamefont {D'Alesio}, \citenamefont {Diehl},\ and\
  \citenamefont {Miller}}]{Bac2004}%
  \BibitemOpen
  \bibfield  {author} {\bibinfo {author} {\bibfnamefont {A.}~\bibnamefont
  {Bacchetta}}, \bibinfo {author} {\bibfnamefont {U.}~\bibnamefont {D'Alesio}},
  \bibinfo {author} {\bibfnamefont {M.}~\bibnamefont {Diehl}}, \ and\ \bibinfo
  {author} {\bibfnamefont {C.~A.}\ \bibnamefont {Miller}},\ }\href {\doibase
  10.1103/PhysRevD.70.117504} {\bibfield  {journal} {\bibinfo  {journal} {Phys.
  Rev.}\ }\textbf {\bibinfo {volume} {D70}},\ \bibinfo {pages} {117504}
  (\bibinfo {year} {2004})}\BibitemShut {NoStop}%
\bibitem [{\citenamefont {Akushevich}\ and\ \citenamefont
  {Shumeiko}(1994)}]{ASh}%
  \BibitemOpen
  \bibfield  {author} {\bibinfo {author} {\bibfnamefont {I.~V.}\ \bibnamefont
  {Akushevich}}\ and\ \bibinfo {author} {\bibfnamefont {N.~M.}\ \bibnamefont
  {Shumeiko}},\ }\href {\doibase 10.1088/0954-3899/20/4/001} {\bibfield
  {journal} {\bibinfo  {journal} {J. Phys.}\ }\textbf {\bibinfo {volume}
  {G20}},\ \bibinfo {pages} {513} (\bibinfo {year} {1994})}\BibitemShut
  {NoStop}%
\bibitem [{\citenamefont {Kotzinian}(1995)}]{aram}%
  \BibitemOpen
  \bibfield  {author} {\bibinfo {author} {\bibfnamefont {A.}~\bibnamefont
  {Kotzinian}},\ }\href {\doibase 10.1016/0550-3213(95)00098-D} {\bibfield
  {journal} {\bibinfo  {journal} {Nucl. Phys.}\ }\textbf {\bibinfo {volume}
  {B441}},\ \bibinfo {pages} {234} (\bibinfo {year} {1995})}\BibitemShut
  {NoStop}%
\bibitem [{\citenamefont {Arens}\ \emph {et~al.}(1997)\citenamefont {Arens},
  \citenamefont {Nachtmann}, \citenamefont {Diehl},\ and\ \citenamefont
  {Landshoff}}]{arens}%
  \BibitemOpen
  \bibfield  {author} {\bibinfo {author} {\bibfnamefont {T.}~\bibnamefont
  {Arens}}, \bibinfo {author} {\bibfnamefont {O.}~\bibnamefont {Nachtmann}},
  \bibinfo {author} {\bibfnamefont {M.}~\bibnamefont {Diehl}}, \ and\ \bibinfo
  {author} {\bibfnamefont {P.~V.}\ \bibnamefont {Landshoff}},\ }\href {\doibase
  10.1007/s002880050430} {\bibfield  {journal} {\bibinfo  {journal} {Z. Phys.}\
  }\textbf {\bibinfo {volume} {C74}},\ \bibinfo {pages} {651} (\bibinfo {year}
  {1997})}\BibitemShut {NoStop}%
\bibitem [{\citenamefont {Bacchetta}\ \emph {et~al.}(2007)\citenamefont
  {Bacchetta}, \citenamefont {Diehl}, \citenamefont {Goeke}, \citenamefont
  {Metz}, \citenamefont {Mulders},\ and\ \citenamefont {Schlegel}}]{bacchetta}%
  \BibitemOpen
  \bibfield  {author} {\bibinfo {author} {\bibfnamefont {A.}~\bibnamefont
  {Bacchetta}}, \bibinfo {author} {\bibfnamefont {M.}~\bibnamefont {Diehl}},
  \bibinfo {author} {\bibfnamefont {K.}~\bibnamefont {Goeke}}, \bibinfo
  {author} {\bibfnamefont {A.}~\bibnamefont {Metz}}, \bibinfo {author}
  {\bibfnamefont {P.~J.}\ \bibnamefont {Mulders}}, \ and\ \bibinfo {author}
  {\bibfnamefont {M.}~\bibnamefont {Schlegel}},\ }\href {\doibase
  10.1088/1126-6708/2007/02/093} {\bibfield  {journal} {\bibinfo  {journal}
  {JHEP}\ }\textbf {\bibinfo {volume} {02}},\ \bibinfo {pages} {093} (\bibinfo
  {year} {2007})}\BibitemShut {NoStop}%
\bibitem [{\citenamefont {Burkhardt}\ and\ \citenamefont
  {Pietrzyk}(1995)}]{vach}%
  \BibitemOpen
  \bibfield  {author} {\bibinfo {author} {\bibfnamefont {H.}~\bibnamefont
  {Burkhardt}}\ and\ \bibinfo {author} {\bibfnamefont {B.}~\bibnamefont
  {Pietrzyk}},\ }\href {\doibase 10.1016/0370-2693(95)00820-B} {\bibfield
  {journal} {\bibinfo  {journal} {Phys. Lett.}\ }\textbf {\bibinfo {volume}
  {B356}},\ \bibinfo {pages} {398} (\bibinfo {year} {1995})}\BibitemShut
  {NoStop}%
\bibitem [{\citenamefont {Akushevich}\ \emph {et~al.}(1998)\citenamefont
  {Akushevich}, \citenamefont {Bottcher},\ and\ \citenamefont
  {Ryckbosch}}]{RADGEN}%
  \BibitemOpen
  \bibfield  {author} {\bibinfo {author} {\bibfnamefont {I.}~\bibnamefont
  {Akushevich}}, \bibinfo {author} {\bibfnamefont {H.}~\bibnamefont
  {Bottcher}}, \ and\ \bibinfo {author} {\bibfnamefont {D.}~\bibnamefont
  {Ryckbosch}},\ }in\ \href@noop {} {\emph {\bibinfo {booktitle} {{Monte Carlo
  generators for HERA physics. Proceedings, Workshop, Hamburg, Germany,
  1998-1999}}}}\ (\bibinfo {year} {1998})\ pp.\ \bibinfo {pages}
  {554--565}\BibitemShut {NoStop}%
\end{thebibliography}%

\end{document}